\documentclass[a4paper,11pt]{spie}
\usepackage{lineno}
\usepackage{hyperref}
\usepackage{upgreek}
\usepackage{tabularx}
\usepackage{mathtools}
\usepackage{enumitem}
\usepackage{amsmath,amssymb}
\usepackage[numbers,square,sort&compress]{natbib}
\usepackage{caption}

\usepackage{booktabs}
\usepackage{multirow}
\usepackage{siunitx} 

\usepackage{float}

\title{In situ cryogenic characterization of proton damage in thick p-channel skipper CCDs}

\author[a]{Brandon M. Roach}
\author[b]{Brenda Cervantes Vergara}
\author[a,b,c]{Alex Drlica-Wagner}
\author[d]{Phoenix Alpine}
\author[e]{Ana Martina Botti}
\author[b]{Claudio Chavez}
\author[c]{Julian Cuevas-Zepeda}
\author[a,b,f]{Juan Estrada}
\author[b]{Guillermo Fernandez Moroni}
\author[g]{Nora Hoch}
\author[h]{Stephen E. Holland}
\author[b,i,j]{Blas Irigoyen Gimenez}
\author[c,j]{Agust{\'i}n Lapi}
\author[b,k,l]{Santiago Perez}
\author[a,b]{Nathan Saffold}
\author[b,k]{Javier Tiffenberg}
\author[m,n]{Yikai Wu}

\affil[a]{\small Kavli Institute for Cosmological Physics, University of Chicago, Chicago, IL, USA}
\affil[b]{\small Fermi National Accelerator Laboratory, Batavia, IL, USA}
\affil[c]{\small Department of Astronomy and Astrophysics, University of Chicago, Chicago, IL, USA}
\affil[d]{\small Department of Aerospace Engineering, University of Illinois Urbana–Champaign, Urbana, IL, USA}
\affil[e]{\small Département de Physique, Université de Montréal, Montréal, Qu{\'e}bec, Canada}
\affil[f]{\small Brookhaven National Laboratory, Upton, NY, USA}
\affil[g]{\small Laboratory for Nuclear Science, Massachusetts Institute of Technology, Cambridge, MA, USA}
\affil[h]{\small Lawrence Berkeley National Laboratory, Berkeley, CA, USA}
\affil[i]{\small Universidad Nacional del Sur, Bah{\'i}a Blanca, Argentina}
\affil[j]{\small Instituto de Investigaciones en Ingenier{\'i}a El{\'e}ctrica ``Alfredo Desages”, CONICET, Bah{\'i}a Blanca, Argentina}
\affil[k]{\small Departamento de F{\'i}sica, Universidad de Buenos Aires, Buenos Aires, Argentina}
\affil[l]{\small Instituto de F{\'i}sica de Buenos Aires, Universidad de Buenos Aires/CONICET, Buenos Aires, Argentina}
\affil[m]{\small Department of Physics and Astronomy, Stony Brook University, Stony Brook, NY, USA}
\affil[n]{\small C.N. Yang Institute for Theoretical Physics, Stony Brook University, Stony Brook, NY, USA}

\authorinfo{Fermilab report number FERMILAB-PUB-26-0399-PPD \\Further author information: \\BMR: roachb@uchicago.edu; BCV:bcervant@fnal.gov;  ADW: kadrlica@fnal.gov}

\begin{document}
\maketitle

\begin{abstract}
Skipper charge-coupled devices (CCDs) are an offshoot of standard silicon pixel detectors and are capable of performing repeated non-destructive charge measurements, enabling deeply sub-electron readout noise. This capability has opened the door to single-photon counting from the near-infrared ($\sim$1.1\,$\upmu$m) to the soft X-ray (several keV), making these devices strong candidates for future astronomical instruments operating in the photon-starved limit. Furthermore, the p-channel architecture used to fabricate Skipper CCDs on n-type silicon has been demonstrated to have an increased hardness to the intense radiation environment of space. Building upon previous irradiation campaigns on room-temperature sensors, here we describe the first radiation-hardness tests of p-channel skipper CCDs at their cryogenic operating temperatures. We assess the performance of the floating-gate output stage and global CCD parameters (charge transfer inefficiency, dark current, hot pixels, and charge traps). We find that these devices maintain excellent performance after displacement damage doses equivalent to ${\sim}$10 years at the Earth/Sun L2 Lagrange point, demonstrating for the first time that these sensors remain radiation-hard in realistic deep-space thermal and radiation environments.
\end{abstract}

\keywords{Skipper CCDs, radiation hardness, displacement damage, pixel detectors, readout noise}

\flushbottom

\section{Introduction}\label{sec:intro}
 
Silicon charge-coupled devices (CCDs) have revolutionized observational astronomy from the X-ray to the near-infrared (NIR; $\lambda \lesssim$1.1\,$\upmu$m) \citep{Boyle:1970, Janesick:2001}. 
Studies of the performance of CCDs in the intense radiation environment of space extends back to the 1970s \citep{Vick:2007}, but became more widely recognized by the astronomical community with the launch of the {\it Hubble Space Telescope (HST)} in 1990 \citep{Blouke:2005}.
Over the last ${\sim}$30 years, there have been numerous studies of the impact of ionizing radiation on CCD detectors both on-orbit \citep[e.g.,][]{ABBEY2003136, STRUDER2003386,Blouke:2005,AMBROSI2002644, Massey:2014ksa, doi:10.2514/1.A36012, 2016A&A...595A...6C, 2026MNRAS.546f2186M,Skottfelt:2024eja,NAGAMATSU2011205,10819429} and in the laboratory \citep[e.g.,][]{polidan2004hot,hopkinson1999,johnson2002analysis,1039641,Dawson:2007yi,GOW201215,8048486,harding2016technology,miyata2003radiation,Gow_2017,Gow_2012}.
Looking ahead, CCDs remain one of the leading technologies under consideration to satisfy the low-noise, high-quantum efficiency, and radiation hardness requirements of the next generation of space-based telescopes. Two of the most demanding scientific objectives for facilities such as the \textit{Habitable Worlds Observatory (HWO)} are the direct imaging and spectroscopy of Earth-mass planets orbiting in the habitable zones of nearby stars~\citep[e.g.,][]{astro2020}. Unfortunately, these planets are extremely faint, with reflected fluxes that are billions of times fainter than their host stars. Even for nearby ($\sim$10 pc) systems, the $V$-band photon flux from such a planet is expected to be ${\sim}10^{-2}$ phot/m$^2$/s~\citep[e.g.,][]{Turyshev:2025pnf}. Even a 6-m-diameter aperture coupled to a perfectly efficient optical path would only give a rate of ${\sim}0.2$ phot/s integrated over the entire focal plane, with the rate per pixel further reduced by point-spread function blurring (in imaging mode) or wavelength dispersion (in spectroscopic mode) to perhaps a few phot/hr. Therefore, the \textit{HWO} mission concept requires UV/VIS/NIR detectors that have extremely low spurious signal rates and the ability to resolve individual photons, while also satisfying the size, weight, power, and thermal requirements of a space telescope \citep[e.g.,][]{2016JATIS...2d1212R,Crill:2022}. 
%
%
\par {At present, one of the limiting factors for conventional CCDs is their read noise $\sigma_r$, which generally manifests as random voltage fluctuations added to each pixel during readout. These fluctuations may arise from many sources with characteristic frequency dependence, including thermal (Johnson-Nyquist) noise, $1/f$ (flicker) noise, and $kTC$ (reset) noise~\citep[e.g.,][]{Janesick:2001}. While some of these noise sources can be mitigated with readout techniques such as correlated double sampling and optimized clocking rates, high-grade astronomical CCDs generally have $\sigma_r \sim \mathrm{\,few\,e}^-$/pixel (rms), limiting their ability to count single charge quanta (and thus single photons). Furthermore, the floating-diffusion output amplifiers of standard CCDs make charge measurement an inherently destructive process: the charge in each pixel can only be measured once before the sense node is reset.
\par The skipper CCD design was proposed in the 1990s, replacing the conventional floating-diffusion output with a floating-gate amplifier structure~\citep[e.g.,][]{10.1117/12.19452,1050535,1990SPIE.1242..238C}, which allows the signal in each pixel to be measured repeatedly and nondestructively by transferring its charge on and off the sense node. By averaging independent measurements (``samples'') of the charge in each pixel, skipper CCDs push below the conventional CCD readout noise floor, achieving
\begin{equation}
    \sigma_r = \frac{\sigma_1}{N_\mathrm{samp}^\alpha},
    \label{eq:readnoise}
\end{equation}
where $\sigma_1$ is the single-sample readout noise (few e$^-$/pixel rms), $N_\mathrm{samp}$ is the number of nondestructive samples per pixel, and $\alpha$ is the noise spectral index (0.5 for an ideal amplifier dominated by white noise). Much of the recent development of skipper CCDs has been performed in the context of rare-event searches such as neutrino and dark-matter detectors~\citep[e.g.,][]{Tiffenberg:2017aac,Cervantes-Vergara:2022ccu,SENSEI:2024yyt,CONNIE:2024off,Alpine:2024kej,DAMIC-M:2025luv}, although recently they have demonstrated single-photon counting at ground-based telescopes~\citep[e.g.,][]{Villalpando:2024oxo}. The latter development has opened new avenues for astronomical observations in the low-signal/low-background regime such as those anticipated for \textit{HWO}.

\par Radiation hardness remains a key area of concern for space-based detectors and microelectronics. High-energy particles (mainly solar and cosmic-ray protons and heavy ions with kinetic energies $\gtrsim$10 MeV/nucleon) can induce a range of damaging effects in semiconductor detectors, causing their performance to degrade over time~\citep[e.g.,][]{Janesick:2001,277547,LINDSTROM200330,2018ITNS...65.1561M}. Ionizing particles can promote charge buildup at the Si/SiO$_2$ interface, increasing leakage current and shifting the voltages delivered to the silicon array~\citep[e.g.,][]{316569,687912,490905,664167}. Additionally, nuclear scattering can displace silicon or dopant atoms from their regular lattice positions, producing charge-trapping sites and increasing charge-transfer inefficiency (CTI) during readout. Since their introduction several decades ago, nearly all scientific CCDs have been n-channel devices (i.e., n-type implanted channels on top of p-type bulk silicon), and this is especially true for space-based\footnote{A notable exception is the Xtend soft X-ray instrument aboard \textit{XRISM}~\citep[e.g.,][]{10.1093/pasj/psaf030}, consisting of four p-channel notch CCDs.} sensors.  Optimizations in the global semiconductor industry make it easier to manufacture n-channel devices~\citep[e.g.,][]{plummer2009silicon}; however, these devices are known to be particularly vulnerable to displacement damage in the space radiation environment. In n-channel devices, the most detrimental radiation-induced traps are phosphorus-vacancy (E-center) and oxygen-vacancy (A-center). These acceptor states readily capture electrons (the majority carriers in n-type silicon) and re-emit them on characteristic timescales ranging from nanoseconds to seconds depending on operating temperature, resulting in charge loss and/or visible smearing of images from astronomical telescopes~\citep[e.g.,][]{doi:10.2514/1.A36012,Massey:2014ksa,2026MNRAS.546f2186M}. In contrast, p-channel CCDs have been found to be significantly more radiation-hard than n-channel devices~\citep[e.g.,][]{1039641,murray2014assessment,Dawson:2007yi,GOW201215,8048486,2013NIMPA.731..160M}. Most critically, p-channel CCDs employ holes as the majority charge-carriers, for which both A-center and E-center defects have much lower capture cross sections compared to electrons. Second, while some E-centers are expected to form in the phosphorus-doped n-type bulk, most p-channel CCDs are optimized for high resistivity (i.e., low bulk phosphorus concentration), meaning that any radiation-induced vacancies will likely combine with each other (divacancy, V$_2$) or oxygen impurities (A-centers) rather than be captured by phosphorus atoms (E-centers). Finally, the ${\sim}$140-K operating temperature of most p-channel CCDs effectively freezes out the remaining defects from emitting dark current. These considerations further motivate the continued testing of p-channel CCDs for space applications.

\par Here, we describe a program to characterize the radiation hardness of thick, p-channel skipper CCDs for future use in space-based instrumentation. This program builds on our previous tests of similar sensors that were irradiated warm and without bias voltages~\cite{Roach:2024iep,Cervantes-Vergara:2025xis,2026arXiv260202461A}. In Section~\ref{sec:setup}, we describe the beamline setup and radiation-damage calculations for our CCDs. In Section~\ref{sec:testplan}, we describe the tests conducted at the Northwestern Medicine Proton Center. In Section~\ref{sec:results}, we describe the results of these tests. In Sec.~\ref{sec:discussion}, we interpret these results in the context of base requirements for facilities like \textit{HWO}. We conclude in Section~\ref{sec:conclusions}.

\section{Experimental setup}\label{sec:setup}

\subsection{CCD detector architecture and packaging}

The sensors tested in this work were ${\sim}650$-$\upmu$m-thick p-channel devices\footnote{As discussed in Sec.~\ref{sec:discussion}, CCDs for space-telescope applications will likely be closer to ${\sim}200\,\upmu\mathrm{m}$ thick, maintaining excellent NIR quantum efficiency while reducing the volume in which displacement damage can occur.} designed by Lawrence Berkeley National Laboratory (LBNL) for the CCD-based dark-matter direct detection experiments SENSEI~\cite{Tiffenberg:2019jde} and Oscura~\cite{2022arXiv220210518A}. These sensors were fabricated on high-resistivity ($\gtrsim 5\,\mathrm{k}\Omega\mathrm{\,cm}$) n-type silicon with $15\times 15\,\upmu\mathrm{m}^2$ pixels.  CCD \#1 had dimensions ($N_\mathrm{row}\times N_\mathrm{col}$) $1278\times 1058$, and CCD \#2 had dimensions $1630\times 588$. Each sensor had four output nodes, one at each corner of the pixel array and containing a floating-gate amplifier that enables repeated nondestructive charge measurement. Each amplifier reads one quadrant of the CCD. In the rest of this paper, we refer to the sensors irradiated in March and July 2025 as CCD \#1 and CCD \#2, respectively. These sensors were from two different vendors (hereafter Vendor \#1 and \#2), and their performance differences are discussed in the following sections. The detector packages were based on those designed for the DAMIC dark-matter detector~\cite{DAMIC:2015ipv} and consisted of the CCD sensor epoxied to a passive silicon wafer (mechanical support) and mounted to a copper tray (thermal interface). The sensors were wirebonded to custom polyimide flexible cables, which provided the electrical connections for input and output voltages. The front side of the package was covered with a 2-mm-thick aluminum sheet to block stray light (visible and infrared) from striking the sensor, with a small hole on one side to allow illumination with a 550-nm green LED. The main departure from the DAMIC design was a square opening in the back of the copper tray, allowing the proton beam to pass through while striking only the aluminum cover and silicon layers, thereby reducing radiogenic activation.

\subsection{CCD test station}\label{sec:test_station}
Initial irradiation studies of p-channel skipper CCDs were performed by bombarding unbiased sensors at room temperature/pressure~\cite{Roach:2024iep,Cervantes-Vergara:2025xis}. Since radiation-damage effects are observed to change with operating temperature and subsequent thermal cycles, we developed a portable vacuum cryostat test station that allowed for irradiation and subsequent characterization of skipper CCDs at their ${\sim}$140--160\,K operating temperature and typical bias voltages, ensuring that any damage was ``frozen'' and not allowed to anneal. The components of the test station are described below and shown in Fig.~\ref{fig:warrenville_setup_wide}.
\begin{itemize}
    \item \textit{Vacuum Cryostat}: The cryostat was a modular $9\times 9\times 9 \mathrm{\,in}^3$ aluminum vacuum chamber. The two faces of the cryostat along the beam axis featured titanium foil windows (6-cm diameter clear aperture and 130-$\upmu\mathrm{m}$ thickness) to minimize inactive material in the beam path while maintaining the vacuum and light-tightness of the cryostat. After exiting the cryostat through the rear titanium window, the proton beam was stopped by a wall of a water-equivalent plastic blocks (``Solid Water''). A turbomolecular pump maintained the residual pressure ${\lesssim}10^{-4}\text{\,mbar}$ at cryogenic temperatures, monitored with a Pirani gauge.
    \item \textit{Cooling and thermal control}: Detector cooling was provided by a mixed-gas Polycold chiller, consisting of a cold end (mounted at the top of the chamber) and compressor. To ensure that the cold end was not struck by protons, it was offset from the top of the chamber with a 6-inch vacuum nipple. Hanging from the base of the cold head was a custom aluminum PCB to which the CCD package was mounted. This PCB included a resistive heater and resistance temperature detector (RTD), allowing closed-loop temperature monitoring and control at the ${<}$0.1-K level with an external Lakeshore temperature controller. The board also featured a central cutout through which the proton beam could safely pass to minimize radiological activation of the aluminum PCB. 
    \item \textit{Detector mounting}: Each detector package was mounted to the bare metal of the aluminum PCB, with a single layer of indium foil serving as a thermal interface. A separate RTD was mounted to the copper tray to monitor the temperature of the CCD package.
    
    \item \textit{CCD interface}: Inside the cryostat, the CCD flexible readout cable connected to a custom vacuum interface board (VIB) developed at Fermilab, which routed the RTD, LED, and CCD connections out of the cryostat while preserving the thermal and vacuum environment.
    Once outside the chamber, the CCD connections passed through a first-stage analog amplification board before arriving at the Low Threshold Acquisition (LTA) controller~\cite{Cancelo:2020egx}. The LTA provided the bias and clock voltages needed to operate the CCD, and also digitized and packaged the CCD video output signals from each readout amplifier into FITS images. A data-acquisition computer running custom software communicated with the LTA over Ethernet. Other boards could also be interposed between the LTA and the VIB to measure and/or inject certain voltages (see, e.g., Secs.~\ref{sec:campaign2} and~\ref{sec:transistors}).
\end{itemize}

\begin{figure}
    \centering
    \includegraphics[width=0.98\linewidth]{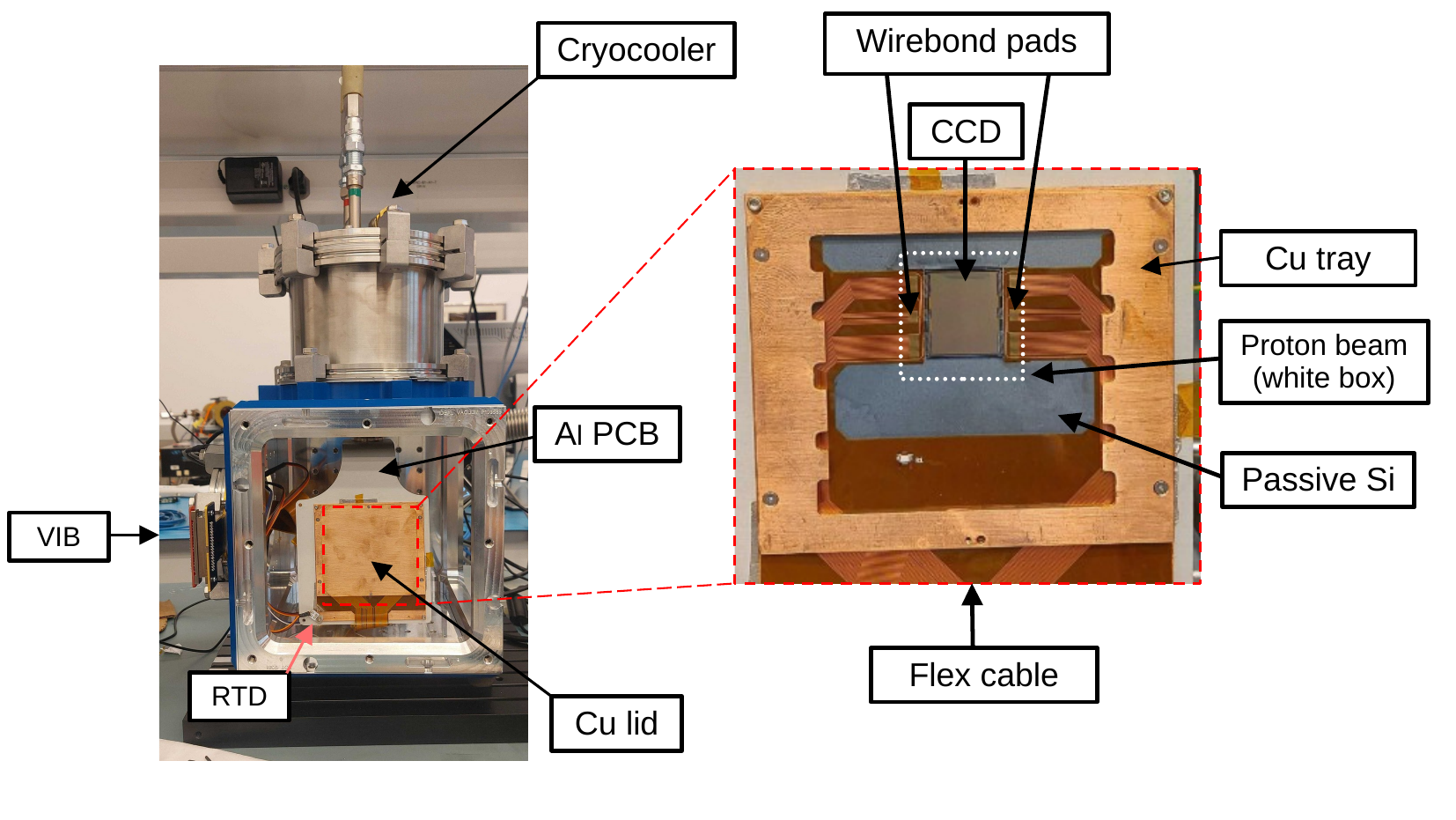}
    \caption{Picture of the CCD characterization and testing system in the clean room at Fermilab, looking down the beam axis. Several sides of the vacuum chamber are removed for an easier view. On the other side of the CCD package, there is a hole in the copper tray and aluminum PCB to allow the proton beam to exit. Prior to irradiation, the copper lid of the sensor package was replaced with a lower-mass aluminum cover to reduce radiogenic activation. For more details, see Sec.~\ref{sec:test_station}.}
    \label{fig:warrenville_setup_interior}
\end{figure}

\begin{figure}
    \centering
    \includegraphics[width=0.98\linewidth]{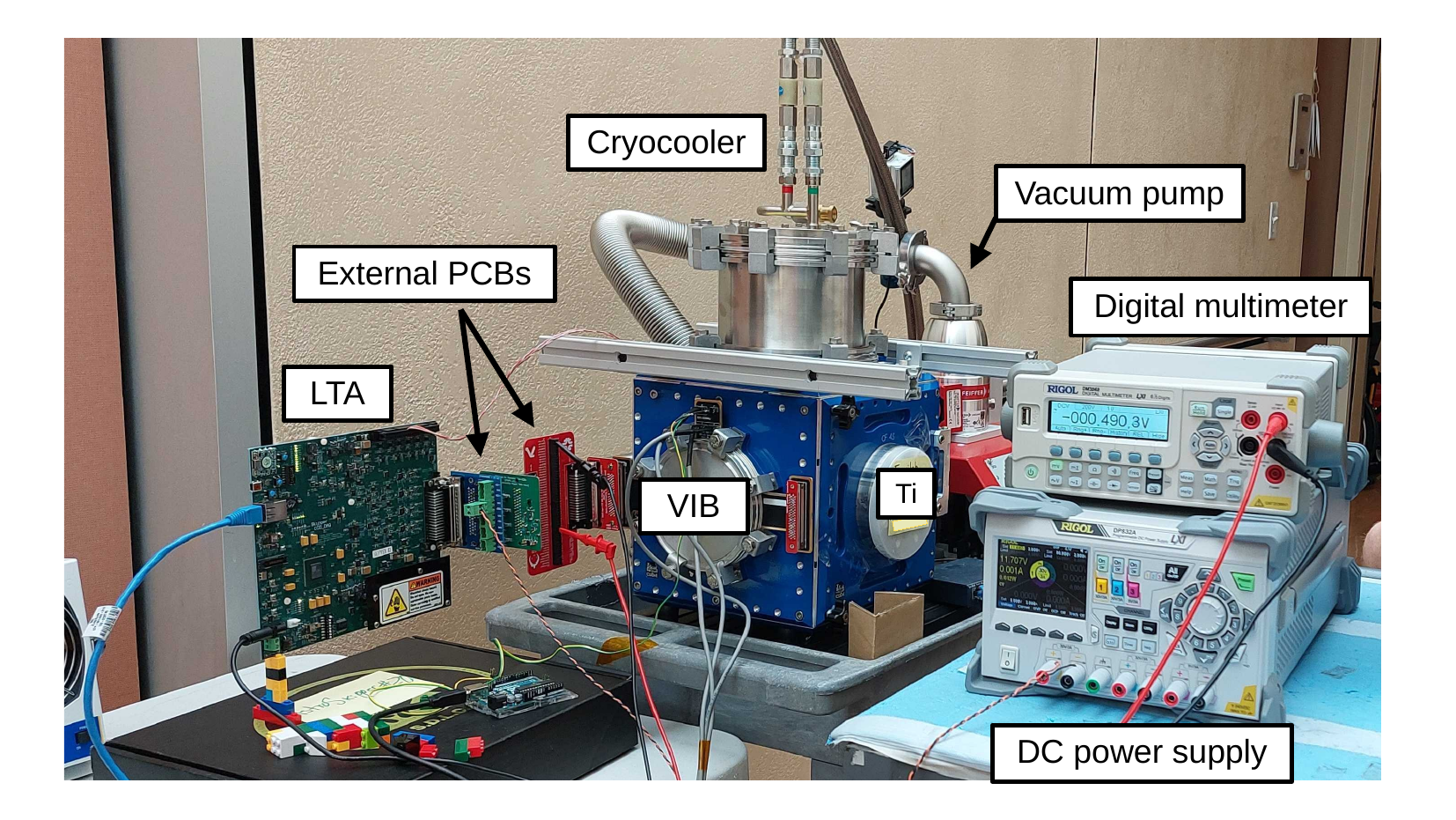}
    \caption{Wide view of the testing system at Northwestern Medicine in an adjacent room following irradiation campaign \#2. All major components are labeled. ``Ti'' refers to one of the titanium foil windows (in this photo covered with a plastic lid for protection). The gray cables from the top of the VIB are the RTD and heater lines, and plug into a Lakeshore temperature controller under the cart (not shown). External PCBs include the first-stage amplification board (green), as well as breakout boards for voltage measurements (red) and injection of an external $V_\mathrm{ref}$ (blue; see Secs.~\ref{sec:campaign2} and ~\ref{sec:transistors}). }
    \label{fig:warrenville_setup_wide}
\end{figure}

\subsection{Northwestern Medicine proton beamline}\label{sec:beamline}
Proton irradiation tests were performed at the Northwestern Medicine Proton Center's IBA C230 cyclotron in Warrenville, IL, USA. Following initial acceleration and extraction from the cyclotron at 230 MeV, the proton beam passed through a tunable low-$Z$ degrader to reduce its maximum energy. A set of magnetic filters and slits was then tuned to select protons with the desired energy of 64 MeV. This energy was chosen to align with previous CCD radiation tests conducted near 60~MeV~\citep[e.g.,][]{jones2000acs,polidan2004hot,Dawson:2007yi} while maintaining the quality and stability of the beam. The outgoing beam shape was defined by a ${\sim}$5-cm-thick brass collimator mounted to the treatment head with a square opening $2\times 2\,\mathrm{cm}^2$. At the CCD position ${\sim}$10 cm from the aperture, the proton flux was approximately uniform across the square area defined by the collimator, and rapidly fell to zero by $2.5\times 2.5\,\mathrm{cm}^2$. The proton flux was continuously monitored during each irradiation test by a thin ionization chamber mounted to the brass collimator, giving a dosimetric uncertainty of a few percent per run. The uncertainty on the proton flux delivered to the CCDs was primarily set by residual uncertainty in the position of the detectors (inside the cryostat) with respect to the beam isocenter. Thus, we assessed an overall 5\% systematic uncertainty on the total fluence. The beam energy was calibrated the day before each run using a dedicated stack of ionization chambers, and the rms energy spread was calculated from the width of the Bragg peak to be $\pm$1.5\% ($\pm1$ MeV for the 64-MeV beam)~\cite{laub}. 
\par To study the particle spectrum at the surface of the CCDs, we constructed a simplified \textsc{Geant4} model using the \texttt{QGSP\_BIC\_HP} physics list, consisting of the brass collimator, titanium entrance window, aluminum module cover, and the silicon CCD itself~\cite{AGOSTINELLI2003250}. We simulated $10^7$ 64-MeV protons uniformly across a $5\times 5$ cm$^2$ beam area (before passing through the collimator), and scored the energies of particles striking the CCD. We found that the attenuated proton energy spectrum was well-described by a Gaussian distribution with mean energy ${\sim}$60 MeV and rms width ${\sim}$0.7 MeV, both of which are completely consistent with analytic estimates from electronic stopping-power tables~\cite{PSTAR}. Combined with the ${\sim}$1-MeV beam-energy rms from the cyclotron itself, the energy spread of the beam at the CCD surface is ${\sim}$1.2 MeV (rms). Furthermore, we find that the losses of protons due to absorption or large-angle scattering in the titanium window or module cover are negligible compared to the few-percent uncertainty on the overall fluence. Finally, we note that high-energy protons can produce secondary neutrons in scattering and spallation reactions, which could contribute to the dose experienced by our sensors. Most of the neutron-generating beamline components (degraders, emittance slits, etc) are out of the direct line of sight from the treatment room, minimizing the neutron fluence. The primary exception was the brass collimator mounted to the treatment head which defined the $2\times 2\mathrm{\,cm}^2$ beam profile. To study the neutron spectrum at the CCDs we used the \textsc{geant4} simulations described above, and found that the neutron fluence at the CCD position is ${\lesssim}0.5\%$ of the proton fluence, with a median kinetic energy of ${\sim}$10 MeV (dominated by forward-peaked direct knock-out reactions in the brass) and a cutoff at ${\sim}$50 MeV. The impact of these neutrons on dose will be discussed in Sec.~\ref{sec:rad_damage_calcs}.


\subsection{Radiation damage calculations}
\label{sec:rad_damage_calcs}
For these tests, we are principally interested in the effects of displacement damage induced by nonionizing energy loss (NIEL)~\citep[e.g.,][]{2007RPPh...70..493L}. Of primary concern is the displacement of atoms from their regular lattice positions, forming vacancies and interstitial defects that may trap charge carriers as they are transferred through the CCD. The displacement damage dose (DDD, MeV/g) delivered by a beam of protons is given by (e.g.,~\cite{summers:1994})
\begin{equation}
\label{eq:ddd}
    \mathrm{DDD} = \int \frac{d\Phi_p}{dE} \; S_\mathrm{nuc}^\mathrm{Si} (E)\,\mathrm{d}E \approx \Phi_p^\mathrm{tot}(E) S_\mathrm{nuc}^\mathrm{Si}(E),
\end{equation}
where $d\Phi_p/dE$ is the differential fluence [p/cm$^2$/MeV], $\Phi_p^\mathrm{tot}$ is the total fluence [p/cm$^2$], $S_\mathrm{nuc}^\mathrm{Si}$ is the nuclear displacement damage kerma in silicon [MeV cm$^2$/g], and the integral is taken over the full kinetic energy range of the proton beam as it passes through the target. The second approximation arises from the assumption that the particle beam is sufficiently monoenergetic (and the sensor sufficiently thin) that variations in $S_\mathrm{nuc}^\mathrm{Si}$ are negligible over the beam energy distribution and the thickness of the device under test. Since the energy lost in the CCD is ${<}$1 MeV and $S_\mathrm{nuc}^\mathrm{Si}(E)$ is slowly varying near 60 MeV, this assumption is well-supported. 
\par To determine the proton fluence targets for our tests, we simulated a simplified mission architecture in SPENVIS~\cite{1999STIN...0021507X,2009EGUGA..11.7457K}, assuming a six-year deployment at the Earth/Sun L2 Lagrange point beginning on 1 January 2026 (i.e., near solar maximum). We used the Emission of Solar Protons (ESP) model to calculate the solar proton fluence at the 95\% confidence level. We then used the SHIELDOSE utility to calculate the DDD experienced at the focal plane from solar protons\footnote{SPENVIS does not include a utility to calculate the shielded DDD contribution from Galactic cosmic rays (GCRs), so following Ref.~\cite{Dawson:2007yi} we include only solar protons in our initial damage calculations. Additionally, the higher energies of GCRs tends to induce spallation reactions in the shielding, further complicating DDD estimates in the absence of a notional shield geometry.}, assuming a silicon detector at the center of an aluminum sphere. Since the geometry of the shielding around the focal plane for an \textit{HWO}-like instrument is unknown, we express the DDD values as a function of aluminum-equivalent absorber thickness---i.e., the thickness of an isotropic aluminum sphere that would yield the same DDD at the focal plane. Previous p-channel CCD irradiation tests targeting the \textit{SNAP} mission assumed an average shielding equivalent to 47 mm aluminum, which would yield an average DDD of ${\sim}6.6\times 10^6$ MeV/g for the six-year mission~\cite{Dawson:2007yi}. Reducing the shielding to 20 (10) mm aluminum equivalent increases the expected focal-plane DDD to ${\sim}2\times 10^7$ (${\sim}4\times 10^7$) MeV/g. We note that the spacecraft body itself would likely provide some solar-proton protection even in the absence of a dedicated focal-plane shield, perhaps equivalent to ${\gtrsim}10$ mm aluminum equivalent~\citep[e.g.,][]{6935036}. To standardize our tests to previous irradiation campaigns, we adopt the 47-mm \textit{SNAP} shielding baseline for these tests, giving an equivalent DDD rate of ${\sim}1.1\times 10^6$ MeV/g/yr at Earth/Sun L2. To evaluate the necessary fluences for the testing at Northwestern Medicine, we scale this DDD rate by the screened relativistic (SR-NIEL) value $S_\mathrm{nuc}^{\mathrm{Si}}(\mathrm{60\,MeV}) \approx 4.0\times 10^{-3} \mathrm{\,MeV\,cm}^{2}/\mathrm{g}$~\cite{sr-niel}, assuming a threshold displacement energy of 21 eV in silicon~\cite{8093101}. This gives an equivalent 1-year fluence $2.7\times 10^8$ p/cm$^2$ for 60-MeV protons, and we use this value to plan our tests. Experimental data on $S_\mathrm{nuc}^\mathrm{Si}$ in this energy range is somewhat sparse, so we assign a 20\% systematic uncertainty on the SR-NIEL value (and thus our required fluences). This reflects the theoretical extrapolation, as well as the uncertainty on the 1-MeV neutron-equivalent damage factor to which the measured $S_\mathrm{nuc}^\mathrm{Si}$ are generally normalized in literature ~\cite{huhtinen_conference}.
\par Though it is largely outside the main scope of the present work, we also briefly describe the total ionizing dose (TID) calculations, which are very similar to Eq.~\eqref{eq:ddd}:
\begin{equation}
    \mathrm{TID} = \int \frac{d\Phi_p}{dE} S_e^\mathrm{Si}(E)\,\mathrm{d}E \approx \Phi_p^\mathrm{tot}(E) S_e^\mathrm{Si}(E).
\end{equation}
The only change is the replacement of the nuclear displacement damage function $S_\mathrm{nuc}^\mathrm{Si}$ with the electronic stopping power $S_e^\mathrm{Si}$ to account for ionization effects in the detector (which can occur in both bulk Si and in the SiO$_2$ gate structures). Assuming an electronic stopping power for 60-MeV protons of $8.6\mathrm{\,MeV\,cm}^2\mathrm{/g}$ for bulk Si (or $9.0\mathrm{\,MeV\,cm}^2\mathrm{/g}$ for SiO$_2$), the TID delivered by an equivalent one-year exposure at L2 is approximately 0.04 krad for both Si and SiO$_2$~\cite{sr-niel}. Again using SPENVIS, we find that the six-year TIDs for aluminum-equivalent shielding thicknesses of 10, 20, and 47 mm are ${\sim}2\,\mathrm{krad}$, ${\sim}1\mathrm{\,krad}$, and ${\sim}0.3\mathrm{\,krad}$, respectively. Therefore, we do not expect the TID effects from our proton-irradiation campaigns to be representative of the space-radiation environment expected at L2, and would thus require a dedicated study which we plan to explore in the future.
\par Finally, we briefly discuss the impact of the fast neutrons produced in the collimator, discussed in Sec.~\ref{sec:beamline}. Again using the SR-NIEL evaluation~\cite{sr-niel}, the stopping power for fast neutrons is in the range $(1\mathrm{-}4)\times 10^{-3}$ MeV cm$^2$/g between 0.1--50 MeV, comparable to the value for 60-MeV protons. Given the factor of ${\gtrsim}10^3$ suppression in fluence between neutrons and protons, we conclude that secondary neutrons have negligible impact on the radiation-hardness constraints we derive in the rest of this paper.

\begin{table}[]
    \centering
    \begin{tabular}{lccc}
        \toprule
        \textbf{Run no.} & \textbf{Fluence} \textbf{[p/cm$^2$]} & \textbf{DDD [MeV/g]} & \textbf{Time at L2 [yr]} \\
        \hline
        \#1 I & $3.8 \times 10^8$ & $1.5\times 10^6$ & 1.4 \\
        \#1 II & $9.3 \times 10^8$ & $3.7 \times 10^6$ & 3.4 \\
        \#1 III & $1.3 \times 10^9$ & $5.3 \times 10^6$ & 4.8 \\
        \#1 IV & $2.7 \times 10^9$ & $1.1 \times 10^7$ & 10.0 \\
        \hline
        \#2 I & $2.8 \times 10^9$ & $1.1 \times 10^7$ & 10.0 \\
        \bottomrule
    
    \end{tabular}
    \caption{List of the irradiation tests performed at Northwestern Medicine with the delivered proton fluence, equivalent DDD at L2, and equivalent time at L2 (assuming 47-mm aluminum shielding equivalent). The fluences for each run of sensor \#1 are cumulative, and all tests for both sensors are at an extracted energy of 64 MeV (reduced to 60 MeV at the CCD). Uncertainties on the delivered fluences are approximately $\pm 5\%$ from ionization-chamber dosimetry and CCD positioning with respect to the beam isocenter. Uncertainties on the DDD are approximately $\pm20\%$ from SR-NIEL extrapolation. See Sec.~\ref{sec:rad_damage_calcs} for further details.}
    \label{tab:test_suite}
\end{table}

\section{Testing Campaigns}\label{sec:testplan}
Proton-irradiation testing occurred in two campaigns, one in March 2025 and the other in July 2025. These test campaigns will be referred to in the remainder of the text as Campaign \#1 and Campaign \#2, respectively.
\subsection{Campaign \#1}
Campaign \#1 was conducted in March 2025, using CCD \#1. As shown in Table~\ref{tab:test_suite}, this campaign was split into four irradiation steps, corresponding to cumulative fluences of $[3.8,9.3,13.3,26.5]\times 10^{8}$ p/cm$^2$. During each irradiation, the sensor was powered off and all CCD pins were grounded together while the sensor was held at a temperature of 160 K. Following each irradiation, the LTA and other readout electronics were reconnected, and the device was subjected to a series of tests, described below. After each image was acquired, the sensor was flushed of charge to reset it for the next exposure.
\begin{itemize}
    \item \textit{Skipper amplifier performance}: Two images were collected with extended serial overscan regions using $N_\mathrm{samp}=300$ to study the performance of the floating-gate output stage. These images only had 50 rows to keep the readout time ${\lesssim}$10 minutes.
    \item \textit{Dark frames}: Six full-frame $N_\mathrm{samp}=1$ images were collected to study any large-scale defects (hot pixels, hot columns, etc) generated by irradiation, as well as the number of charged-particle tracks emitted by radiogenic activation products (Fig.~\ref{fig:microchip_betas}).
    \item \textit{Flat fields}: Three $N_\mathrm{samp}=300$ images were collected with an LED uniformly illuminating the sensor to study the effects of charge-transfer inefficiency (CTI). The average pixel charge occupancy was ${\sim}10^3$ e$^-$, which was chosen to allow easy comparisons with X-ray hits produced during the decay of $^{55}\mathrm{Fe}$ (${\sim}$1600 e$^-$ for single-pixel events). The image dimensions were the same as the ``Skipper amplifier performance'' images.

\end{itemize}
Following the final irradiation, the entire testing system was moved to an adjacent storage room while maintaining a temperature of ${<}$170~K. The sensor was then powered back on and the aforementioned tests were repeated for the next four days to track any long-term changes in performance after irradiation. Following this period, the cryostat was slowly brought back to room temperature (in 10-K steps from 160--200 K) to study the evolution of the dark current and charge-trap population. The CCD package was then returned to Fermilab for long-term storage. 

\subsection{Campaign \#2}\label{sec:campaign2}
\par Campaign \#2 was conducted in July 2025. Most key elements stayed the same (beam energy, thermal/vacuum chamber design, etc), but there were several key differences, which we describe here. Perhaps the most critical difference between Test \#1 and Test \#2 was the CCD itself. As will be discussed in the following sections, previous studies of CCDs from Vendor \#1 have found a population of charge-trapping states consistent with transition-metal impurities~\cite{PerezGarcia:2024lqj, Cervantes-Vergara:2025xis}. This leads to an overall increase in CTI and complicates the analysis of lattice defects generated from proton exposure. For Test \#2, we packaged a sensor from another supplier (Vendor \#2) known to have a much lower rate of these impurity traps~\cite{Cervantes-Vergara:2025xis}. This was confirmed by pre-irradiation measurements of CTI and dark current. Owing to time constraints at the beamline, we chose a single fluence target of $2.7\times 10^9$ p/cm$^2$ at an extracted beam energy of 64 MeV. We also wrapped the sensor package and aluminum PCB in multilayer insulation (MLI), consisting of ${\sim}10$ layers of aluminized Mylar film and fabric mesh. This blocked both stray light and infrared radiation emitted from the warm interior walls of the cryostat, minimizing the dark count rate in the CCD and allowing us to lower the operating temperature to 140 K. The energy lost by the protons in the MLI (total thickness $<$1 mm) was negligible compared to the other passive components in the beam path.
\par To ensure the sensor could actively operate in a realistic radiation environment, we left the LTA controller and other readout electronics connected during the beam exposure. All bias and clocking voltages were set to their nominal operational values, and the sensor was commanded to continuously flush charge during the irradiation at a rate of several hundred kHz to minimize charge buildup in the bulk. Following irradiation, the sensor was powered down and the readout electronics and cryocooler were temporarily disconnected for the quick move to the adjacent testing room. The chamber remained under vacuum, and the maximum temperature of the CCD was ${\sim}$155 K before being slowly returned to 140 K. 
\par The post-irradiation tests performed on Sensor \#2 were generally identical to Sensor \#1, with one new addition. The output stage of each skipper amplifier contains a p-type MOSFET (M1) which can experience TID-related damage due to charging of the gate oxides. This generally manifests as shifts in the characteristic curves of M1---i.e., $I_{ds}$ versus $V_{gs}$ ($=V_g - V_s$) or $V_{ds}$ ($=V_{dd} - V_s)$. To probe any voltage shifts on the output node, we adopted the same general procedure as Ref.~\cite{Cervantes-Vergara:2025xis}. We used an external power supply and a breakout board to force the reset MOSFET into conduction mode, thus setting $V_g = V_\mathrm{ref}$ in M1. All other bias voltages (with the same values as the other tests) were supplied and monitored with the LTA. The results of these transistor-curve scans are discussed in Sec.~\ref{sec:transistors}.
\par Approximately two hours after the transistor-curve scans were finished, the entire system was moved to an adjacent holding room where all tests (with the exception of the transistor-curve measurements) would continue to run on a loop to study long-term evolution of the sensor. The sensor remained below 160 K during this move, and was allowed to stabilize at 140 K before resuming the tests. Similarly to CCD~\#1, five days after irradiation we initiated a controlled warm-up of the sensor to study dark current in 10-K intervals. At the conclusion of these tests, the system was returned to room temperature and brought back to Fermilab, where the detector was placed in room-temperature storage.

\begin{figure}
    \centering
    \includegraphics[width=0.49\linewidth]{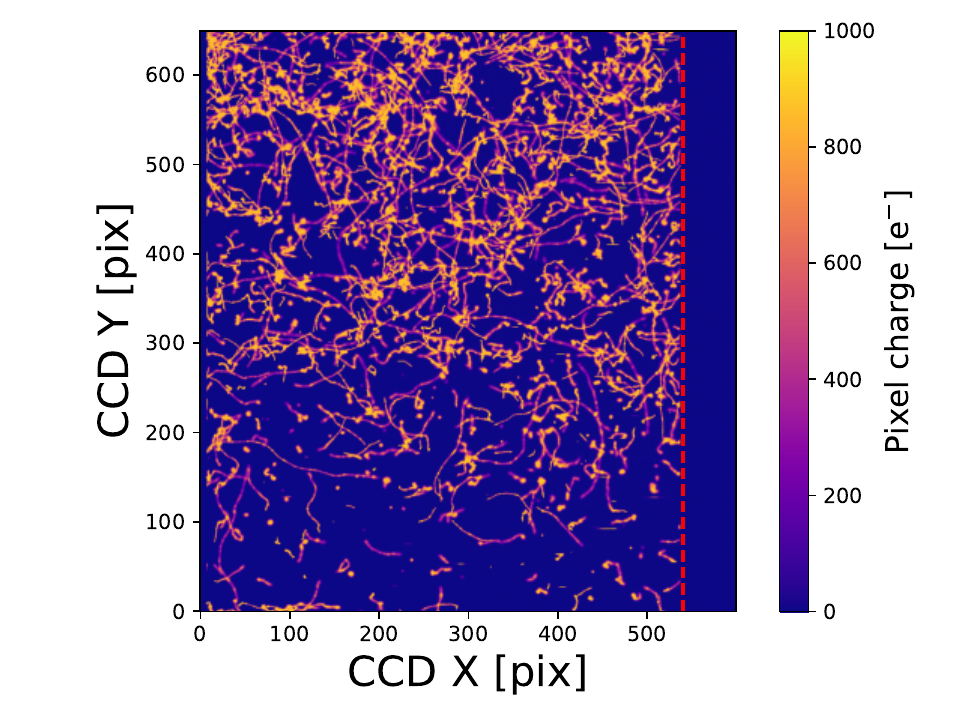}
    \includegraphics[width=0.49\linewidth]{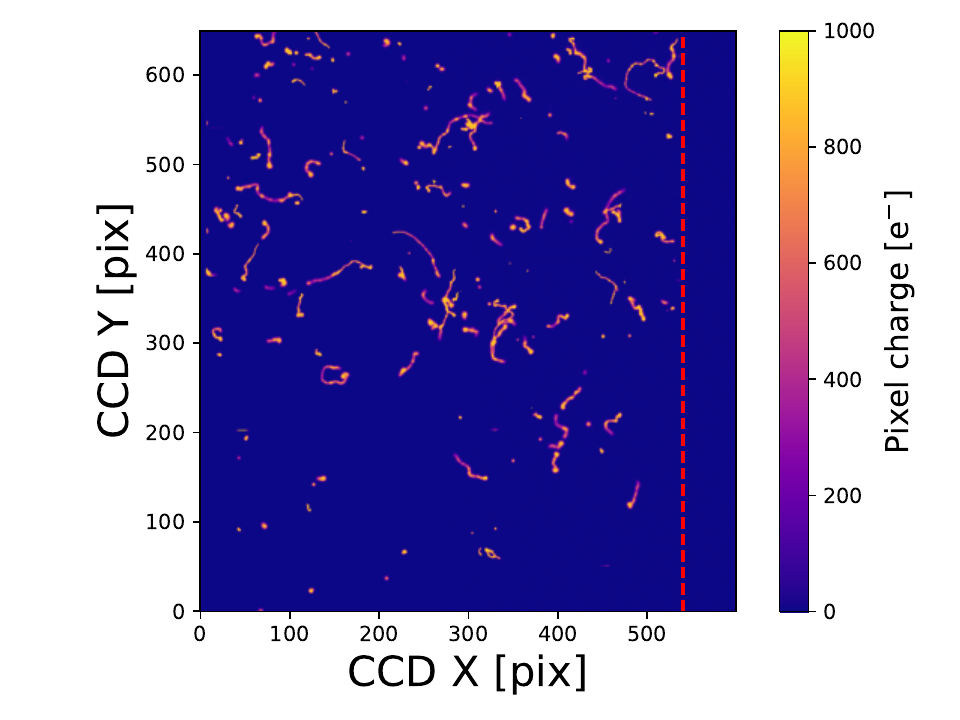}

    \caption{Single-sample images recorded by one quadrant of CCD \#1 approximately 10 minutes (left) and 1 hour (right) after receiving its final dose of $1.3\times 10^9$ p/cm$^2$. Significant numbers of charged-particle tracks are visible, mainly $\beta^\pm$-particles emitted by unstable nuclei and Compton-scattered silicon electrons initiated by $\gamma$-ray emitters. The dashed vertical line indicates the start of the serial overscan region. The equivalent exposure time per pixel increases vertically from 0 s to approximately 80 s (the readout time per frame).}
    \label{fig:microchip_betas}
\end{figure}

\section{Results}\label{sec:results}

\subsection{Skipper amplifier performance}\label{sec:amp_performance}
The floating-gate output stage (``skipper amplifier'') is the main architectural difference between skipper CCDs and conventional p-channel CCDs, and thus one of the central goals of our irradiation campaign is to verify its radiation-hardness. We consider two primary observables: the scaling of readout noise $\sigma_r$ [$\mathrm{e}^-$ rms/pixel] with $N_\mathrm{samp}$, and the stability of the amplifier gain $g_e$ [ADU/e$^-$].
\par To calculate these quantities, we use the deep multi-sample datasets described in Sec.~\ref{sec:testplan}. Each amplifier produces a ``data cube'' with dimensions $(N_\mathrm{rows},N_\mathrm{cols},N_\mathrm{samp})$, where each of the $N_\mathrm{samp}$ slices of $(N_\mathrm{rows},N_\mathrm{cols})$ are independent nondestructive samples of the pixel array read by the corresponding skipper amplifier. We construct an image $\bar{A}$ whose pixel values in ADU are the running average of all previous samples, i.e.,
\begin{equation}
    \bar{A}_{ij}(N_\mathrm{samp}) = \frac{1}{N_\mathrm{samp}} \sum_{k = 1}^{N_\mathrm{samp}} A_{ij}^{(k)},
    \label{eq:skipper_average_matrix}
\end{equation}
where the superscript $(k)$ denotes the sample index. To determine $g_e$ and $\sigma_r$, we collapse the array in Eq.~\eqref{eq:skipper_average_matrix} from each readout amplifier into a set of $N_\mathrm{samp}$ one-dimensional histograms to which we fit a comb of Gaussian functions representing the quantized electron peaks. Owing to the low-signal linearity of the sensors, we require that each peak be spaced a distance $g_e$ from its neighbors, with all peaks having the same $\sigma_r$. Furthermore, since only $\sigma_r$ scales with $N_\mathrm{samp}$, we freeze the model for each amplifier to the best-fit values of the peak centroids and $g_e$ obtained at the maximum value of $N_\mathrm{samp}$, and only allow $\sigma_r$ to vary. Finally, for each amplifier we perform a least-squares fit of the $(N_\mathrm{samp},\sigma_r)$ curves to the generalized skipper CCD noise equation in Eq.~\eqref{eq:readnoise} to extract the noise spectral index $\alpha$.
\par The results of this analysis are shown in Table~\ref{tab:ccd_amps_detailed} and an example is shown in Fig.~\ref{fig:readoutnoise}. For both CCDs tested, we find that most skipper amplifiers continue to perform nominally, even up to the highest fluences tested. We observe small inter-test variations in $\sigma_1$ and $g_e$, which we attribute to disconnecting and reconnecting the external readout electronics. We observe no measurable change in $\sigma_1$ and $g_e$ with fluence, which is consistent with the small TID (${<}1\mathrm{\,krad}$) delivered to the sensor. Furthermore, the noise spectral index is $0.45\lesssim \alpha \lesssim 0.55$ for almost all amplifiers, indicating near-ideal $1/\sqrt{N_\mathrm{samp}}$ noise scaling with only minimal $1/f$ noise leakage. The unusual noise characteristics of amplifiers A and B on CCD \#2 are still under investigation. (We note that these two amplifiers exhibit significant clock-induced charge prior to irradiation, suggesting their atypical behavior is not a reuslt of radiation damage.) Taken together, these results are consistent with previous tests of irradiated p-channel skipper CCDs~\cite{Roach:2024iep,Cervantes-Vergara:2025xis} here extended to more realistic (cold and biased) operating conditions.

\begin{figure}
    \centering

    \includegraphics[width=0.475\linewidth]{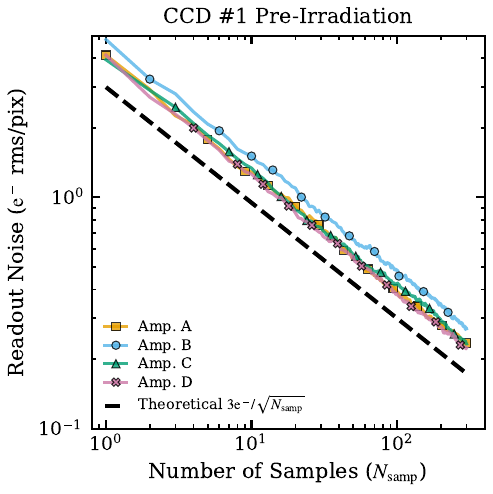}
    \includegraphics[width=0.475\linewidth]{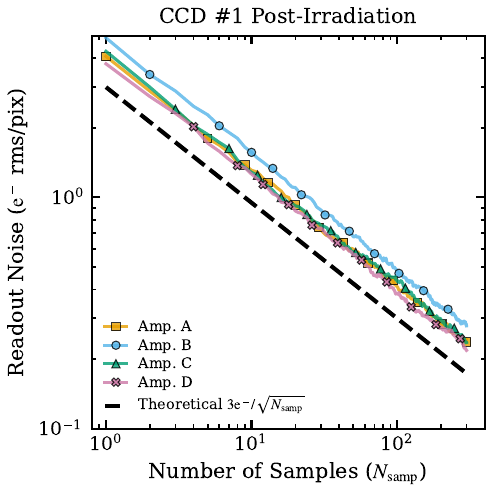}

    \vspace{0.5cm}
    \includegraphics[width=0.475\linewidth]{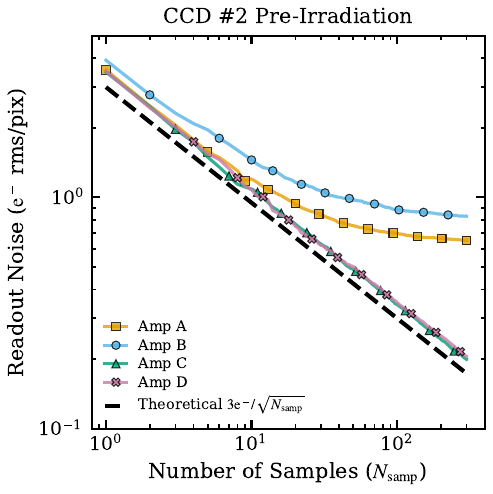}
    \includegraphics[width=0.475\linewidth]{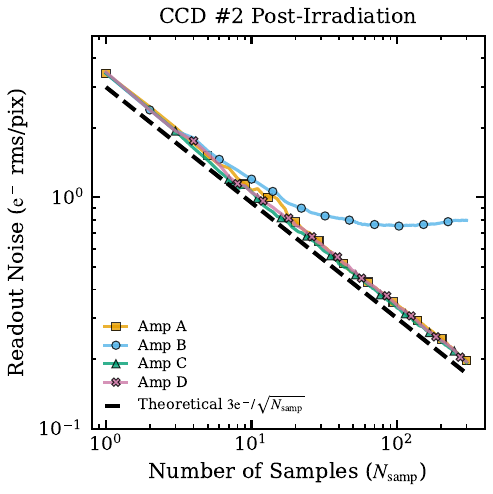}
    \caption{Plots of readout noise $\sigma$ versus $N_\mathrm{samp}$ for CCD \#1 (left) and CCD \#2 (right), as measured at Northwestern Medicine several days after irradiation. The dashed line showing the expected scaling for an ideal skipper CCD with $\sigma_1 = 3\mathrm{\,e}^-$ rms/pix is included to guide the eye. Most amplifiers show nominal $1/\sqrt{N_\mathrm{samp}}$ scaling both before and after an equivalent 10-year DDD, except for amplifiers A and B on CCD~\#2.}
    \label{fig:readoutnoise}
\end{figure}

\subsection{Charge Transfer Inefficiency}\label{sec:cti}
\par Charge transfer inefficiency (CTI) measures the fractional charge loss per transfer as the charge packet is clocked between CCD pixels. CTI is the result of the complex interplay between clocking voltages, operating temperature, and the behavior of charge traps~\citep[e.g.,][]{Janesick:2001}. In the Shockley-Read-Hall (SRH) model of trap kinetics, the emission time constant of a charge trap in silicon is given by~\citep[e.g.,][]{PhysRev.87.835,Cervantes-Vergara:2025xis}
\begin{equation}
    \begin{dcases}
    \tau_e = \frac{1}{\sigma v_\mathrm{th} N_v} \mathrm{exp}\left[\frac{E_a}{kT}\right]\\
    v_\mathrm{th} = \sqrt{\frac{3kT}{m_\mathrm{cond}^{(h)}}} \\
    N_v = 2\left(\frac{2\pi kT m_\mathrm{dens}^{(h)}}{h^2}\right)^{3/2}.
    \end{dcases}
    \label{eq:srh}
\end{equation}
Here $\sigma$ is the trap cross section; $T$ is the temperature; $E_a$ is the energy of the trap above the valence level; $m_\mathrm{cond}^{(h)} \approx 0.41 m_e$ and $m_\mathrm{dens}^{(h)}\approx 0.94 m_e$ are the effective hole $(h)$ masses for conductivity and density-of-states calculations, respectively~\cite{10.1063/1.345414}; $m_e$ is the electron rest mass; $k$ is Boltzmann's constant; and $h$ (without parentheses) is Planck's constant. In particular, if the trap emission time constant $\tau_e \gtrsim \Delta t$ (where $\Delta t$ is the inter-pixel transfer time), then any charge trapped will tend to be re-emitted into later pixels, manifesting as an increase in CTI. The CTI of a CCD is generally characterized in both the serial (horizontal) and parallel (vertical) directions, owing to their different origins and timescales as a consequence of the CCD readout sequence. Serial CTI can only occur in the serial-register row, whereas parallel CTI can occur anywhere in the active area of the sensor.

\par We measured CTI using the extended pixel edge response (EPER) method rather than using a radioactive source such as $^{55}\mathrm{Fe}$ ~\citep[e.g.,][]{Janesick:2001}. For these tests, the sensor was uniformly illuminated with a green LED to a signal level of ${\sim}10^3\mathrm{\,e}^-/\mathrm{pix}$ to allow for comparison with other CCDs tested using $^{55}\mathrm{Fe}$. The serial and parallel CTI are calculated using the equations
\begin{equation}
    \mathrm{CTI}_\mathrm{ser} = \frac{ Q_\mathrm{OS}}{N_\mathrm{ser} Q_\mathrm{act}}\;\;\;\;\;\;\mathrm{and}\;\;\;\;\ \mathrm{CTI}_\mathrm{par} = \frac{Q_\mathrm{OS}}{N_\mathrm{par} Q_\mathrm{act}}
\end{equation}
where $Q_\mathrm{OS}$ is the total charge in the overscan, $N_\mathrm{ser}$ ($N_\mathrm{par}$) is the number of serial (parallel) pixel transfers to reach the sense node, and $Q_\mathrm{act}$ is the charge in the last few pixels of the active area before the overscan. To convert between pixel position and time, we use the relations
\begin{equation}  
    \begin{dcases}
        \Delta t_\mathrm{ser} \simeq \Delta t_\mathrm{clock,ser} + N_\mathrm{samp}\Delta t_\mathrm{samp} \\
        \Delta t_\mathrm{par} \simeq \Delta t_\mathrm{row} = \Delta t_\mathrm{clock,par} + (N_\mathrm{col,CCD}+N_\mathrm{col,OS})\Delta t_\mathrm{ser}
    \end{dcases}
\end{equation}
Here, $\Delta t_\mathrm{clock,*}$ is the time required to shift charge by one pixel in the corresponding direction, $\Delta t_\mathrm{samp}$ is the time required to perform a single-sample measurement of the pixel charge, and $\Delta t_\mathrm{line}$ is the time required to read out an entire row including both physical (CCD) and overscan (OS) pixels. For both sensors, we used the same clocking parameters: $\Delta t_\mathrm{clock,ser} = 27.8\,\upmu\mathrm{s}$, $\Delta t_\mathrm{samp} = 50.9\,\upmu\mathrm{s}$, and $\Delta t_\mathrm{clock,par} = 40.0\,\upmu\mathrm{s}$. The corresponding readout timescales are shown in Table~\ref{tab:cti}. We then fit a function of the form
\begin{equation}
    \mathcal{Q}(t) = C_1 \mathrm{exp}\left[-\frac{t}{\tau_e}\right] + C_\mathrm{2} \mathrm{exp}\left[-\frac{t}{\tau_\mathrm{RC}}\right] + C_\mathrm{3}x + C_4
\end{equation}
to the one-dimensional overscan profiles to model the charge $\mathcal{Q}$ as a function of time. The $C_1$ term corresponds to emission from the dominant trap species with time constant $\tau_e$. The $C_2$ term is used to model the large negative $RC$ transient in the $N_\mathrm{samp}=1$ serial overscan of CCD~\#2 as the readout electronics transition from the large signal levels in the active area to the low signal levels in the overscan. The $C_\mathrm{3}$ and $C_\mathrm{4}$ terms account for any residual linear trend in the baseline (e.g., as a result of clock-induced charge). Leaving the four $C_i$ and the $\tau_e$ as free parameters to fit, the total deferred charge is then simply $Q_\mathrm{OS} = \int_0^\infty C_1 \exp[-t/\tau_e]\,\mathrm{d}t$. We note that this functional form assumes only a single trap species dominates the CTI in each readout configuration, though owing to the difference in timescales between the relevant trap species and the few-hundred-pixel lengths of the overscan regions, this assumption is reasonable.
\par Before discussing the results of these irradiation campaigns, we briefly review the most significant charge-trapping species in our thick p-channel CCDs. Previous pocket-pumping studies of similar devices~\citep[e.g.,][]{Cervantes-Vergara:2025xis,PerezGarcia:2024lqj} found four major trapping states, summarized below:
\begin{itemize}
    \item \textit{Divacancy donor level} (hereafter simply \textit{divacancy} or $\mathrm{V}_2$) traps have energies ${\sim}$0.18--0.23\,eV above the valence level, and cross sections $\sigma \sim (0.1\mathrm{-}1)\times 10^{-15}\,\mathrm{cm}^2$. Previous studies favored the upper end of both ranges, suggesting $\tau_e \sim \mathrm{\,few\,ms}$ at 140~K and ${\sim}200\,\upmu\mathrm{s}$ at 160~K.
    \item \textit{Vacancy/oxygen complexes} (V$_n$O$_m$) have been observed at ${\sim}$0.25 eV above the valence band with $\sigma \sim 2\times 10^{-14}$. This corresponds to $\tau_e \sim 1\mathrm{\,ms}$ at 140~K and ${\sim}100\,\upmu\mathrm{s}$ at 160~K.
    \item \textit{Transition-metal impurities} introduced during fabrication span a range of energies~\citep[e.g.,][]{metalimpurities}, but previous studies suggest energies ${\sim}$0.34~eV above the valence band and $\sigma \sim 3\times 10^{-15}\mathrm{\,cm}^2$. These imply $\tau_e \sim 10\mathrm{\,s}$ at 140~K and ${\sim}200\mathrm{\,ms}$ at 160~K. 
    \item \textit{Carbon/oxygen interstitial} (C$_i$O$_i$) traps have energies ${\sim}0.39\mathrm{\,eV}$ above the valence band, and $\sigma \sim 10^{-14}\,\mathrm{cm}^2$. This implies $\tau_e \sim 10\mathrm{\,min}$ at 140~K and ${\sim}5\mathrm{\,s}$ at 160~K, both of which are far too long to observe with the present EPER measurements.
\end{itemize}
\begin{figure}[b]
\begin{minipage}{\textwidth}
{\centering

        \small 
    \setlength{\tabcolsep}{4pt} 
    \renewcommand{\arraystretch}{1.2}

    \begin{tabular}{@{} ll cc cccc @{}} 
        \toprule
        & & \multicolumn{2}{c}{\textbf{Timescale}} & \multicolumn{2}{c}{\textbf{Pre-Rad.}} & \multicolumn{2}{c}{\textbf{Post-Rad.}} \\
        \cmidrule(lr){3-4} \cmidrule(lr){5-6} \cmidrule(lr){7-8}
        \textbf{Sensor} & $\mathbf{N_{samp}}$ & $\Delta t_{\text{ser}}$ & $\Delta t_{\text{par}}$ & \textbf{Serial} & \textbf{Parallel} & \textbf{Serial} & \textbf{Parallel} \\
        \midrule
        
        \multirow{2}{*}{CCD \#1} 
        & 1   & ${\sim} 79\,\upmu$s & ${\sim} 47$\,ms & $<10^{-5}$ & (1.9--2.9)$\times 10^{-4}$ & (1.8--3.5)$ \times 10^{-5}$ & (1.5--2.2)$ \times 10^{-4}$ \\
        & 300 & ${\sim} 15$\,ms    & ${\sim} 11$\,s   & (1.0--2.0)$ \times 10^{-4}$ & N/A & (0.5--1.3)$ \times 10^{-4}$ & N/A \\
        \addlinespace
        
        \multirow{2}{*}{CCD \#2} 
        & 1   & ${\sim} 79\,\upmu$s & ${\sim} 47$\,ms & $< 10^{-6}$ & $< 10^{-6}$ & (1.0--1.6)$\times 10^{-6}$ & $< 10^{-6}$ \\
        & 300 & ${\sim} 15$\,ms    & ${\sim} 11$\,s   & $< 10^{-6}$ & N/A & (1.0--1.5)$\times 10^{-6}$ & N/A \\
        \bottomrule
    \end{tabular}        

}

\end{minipage}
        \captionof{table}{Extracted Charge Transfer Inefficiency (CTI) in the serial and parallel directions for CCD \#1 and CCD \#2, tested at 160~K and 140~K, respectively, following the full proton dose. Note that the $N_\mathrm{samp}=300$ parallel CTI was not measured due to time constraints. For more details, see Sec.~\ref{sec:cti}.}

    \label{tab:cti}
\end{figure}
\begin{figure*}[htpb]
\includegraphics[width=\textwidth]{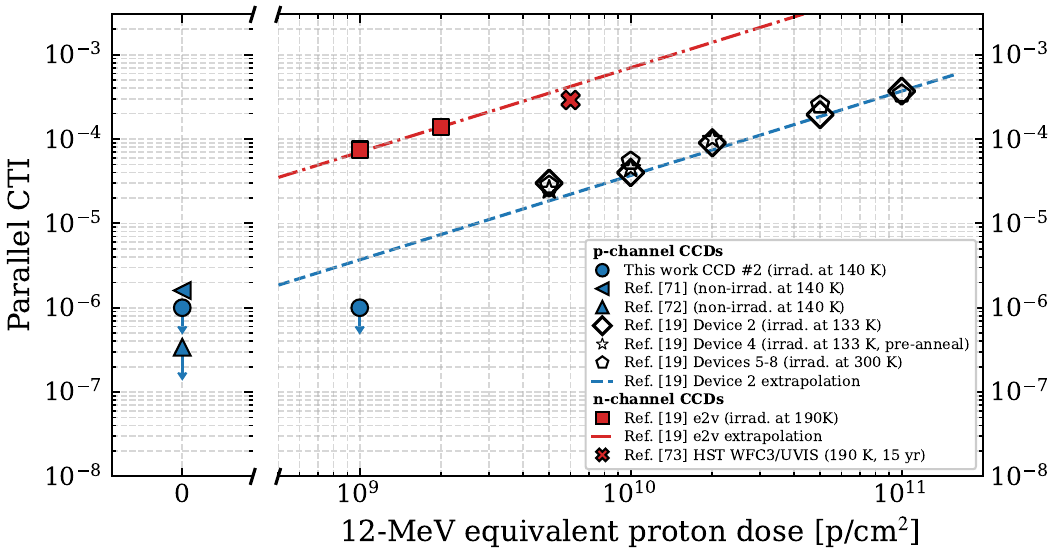}
\includegraphics[width=\textwidth]{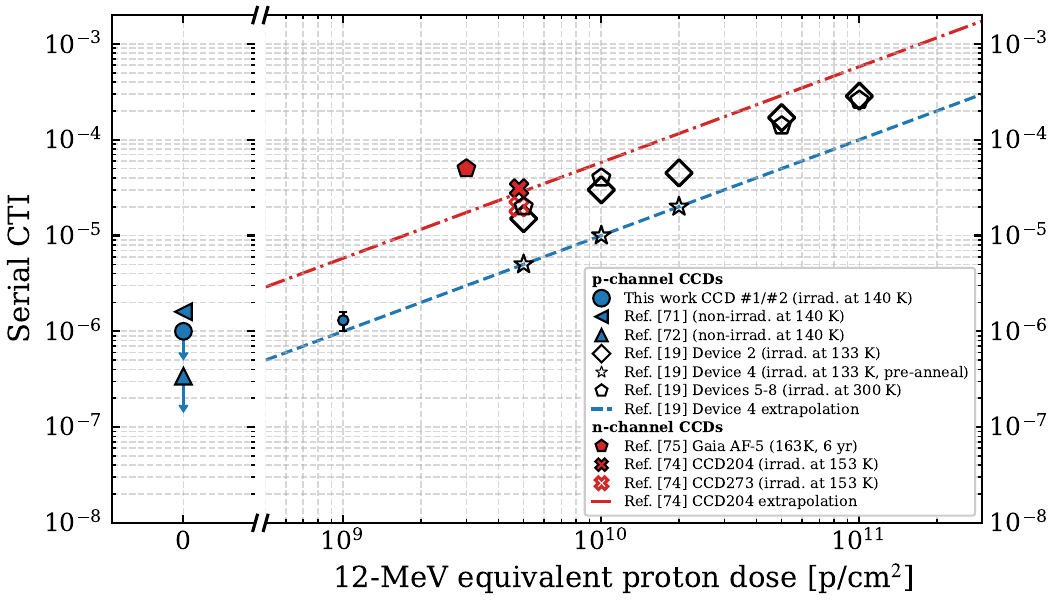}
\caption{\textbf{(Top)} Parallel CTI (measured under a range of temperatures and conditions) for p-channel and n-channel CCDs before and after irradiation. Irradiation has been normalized to the 12-MeV equivalent proton dose following Ref.~\cite{Dawson:2007yi}. For CCD \#2, the data are for $N_\mathrm{samp}=1$ only. Additional p-channel skipper CCDs from Vendor \#2 are described in Refs.~\cite{2020SPIE11454E..1AD,2024PASP..136d5001M}.  CCDs from Vendor \#1 are not shown, owing to their large pre-irradiation CTI from fabrication-induced traps. The \textit{HST} WFC3/UVIS data are from Ref.~\cite{oconnor2025wfc3}. \textbf{(Bottom)} Same as above, but for the serial CTI. Note that the points for CCD \#1/\#2 lie on top of each other). The \textit{Euclid}-like CCD204 and CCD273 data are from Ref.~\cite{2012SPIE.8453E..16G}, and the \textit{Gaia} data (quoted for a signal level of $5\,000$ e$^-$/pix) are from Ref.~\cite{2022JATIS...8a6003A}.}
\label{fig:parallelcti}
\end{figure*}
\pagebreak
\par For CCD \#1, we observe significant CTI both pre- and post-irradiation. Our $N_\mathrm{samp}=300$ serial CTI data and $N_\mathrm{samp}=1$ parallel CTI data both show clear exponential signals with time constants ${\sim}3\mathrm{\,pix}$ and ${\sim}1\mathrm{\,pix}$, respectively. Correcting for the pixel clocking rates implies $\tau_e$ in the range ${\sim}$50--75~ms, suggesting that the poor CTI is dominated by a common trap species. Since the magnitude of the CTI remained unchanged after the 10-year DDD, the trap species is likely not related to the proton bombardment. At the 160-K operating temperature, the transition-metal impurities with $\tau_e \sim 200\mathrm{\,ms}$~\citep[e.g.,][]{PerezGarcia:2024lqj,Cervantes-Vergara:2025xis} are the obvious candidate, especially considering minor variations in operating temperature. Similarly, the $N_\mathrm{samp}=1$ serial CTI indicates a trapping species with $\tau_e \sim 5\mathrm{\,pix} \sim 75\,\upmu\mathrm{s}$ that is formed at small-yet-detectable levels during irradiation, consistent with a $\mathrm{V}_2$ defect. We note that our pre-irradiation CTI measurements are consistent with Ref.~\cite{2026arXiv260202461A} which characterized a sensor from the same vendor with $^{55}\mathrm{Fe}$ X-rays using $N_\mathrm{samp}=10$ (i.e., $\Delta t_\mathrm{par} \sim \mathrm{few\;hundred\;ms}$) at 153~K. That study found a similar non-irradiated parallel CTI of ${\sim}$(1.1--2.5)$\times 10^{-4}$, indicating that both sensors were affected by the same pre-irradiation impurities and supporting the robustness of our EPER measurements. 
\par For CCD \#2 we find significantly smaller CTI across all testing points, indicating an exceptionally small density of traps with $\tau_e$ between ${\sim}$0.1--100\,ms. In all cases shown in Table~\ref{tab:cti} the CTI is ${\lesssim}1.5\times 10^{-6}$, and there are no exponential trails visible in either the serial or parallel overscan. This indicates that the trap formation rate must be extremely small, and/or that any traps formed must have time constants either much shorter or much longer than our CCD clocking timescales. The apparent lack of the V$_2$ trap in these data may be explained by the 140~K temperature at which our device was irradiated. First, the V$_2$ has $\tau_e \sim \mathrm{few\,s}$ at this temperature, which is at least an order of magnitude longer than $\Delta t_\mathrm{ser}$ at $N_\mathrm{samp}=1$; thus, any trapped charge would be emitted across a large stretch of overscan, hindering detection. Additionally, positive vacancies in p-type silicon only become mobile above ${\sim}$150~K; thus, CCD \#2 being irradiated and tested at a lower temperature may have suppressed V$_2$ trap formation. In contrast, p-channel devices irradiated at warmer temperatures (e.g., 153~K in Ref.~\cite{8048486}, 160~K for CCD \#1, and 300~K in Ref.~\cite{Cervantes-Vergara:2025xis}) suggest a much larger V$_2$ concentration, owing to the greater mobility of vacancies at elevated temperatures. We do not expect the C/O trap complexes to be visible in any of our EPER measurements for this device, since their $\tau_e \sim 10\mathrm{\,min}$ at these low temperatures. 
\par We conclude the discussion of CTI by comparing these results with previous irradiation tests of both p- and n-channel CCDs. The results are shown in Fig.~\ref{fig:parallelcti}, demonstrating the  advantage of p-channel CCDs vis-{\`a}-vis reducing displacement-damage-induced CTI. For parallel CTI, our results for CCD~\#2 (which was not afflicted with pre-existing fabrication traps) meet or exceed the expected CTI damage curves extrapolated from Ref.~\cite{Dawson:2007yi} calibrated using $^{55}\mathrm{Fe}$, for both single-sample and deeply-sub-electron configurations. Though our EPER measurements are most sensitive to traps with $\tau_e \sim \Delta t$ (rendering long-lived traps such as C$_i$O$_i$ invisible), we note that our CTI measurements are still in excellent agreement with the aforementioned $^{55}\mathrm{Fe}$ measurements. Further tests of irradiated p-channel skipper CCDs maintained at cryogenic temperatures and searching for CTI in the ``disappearance channel'' (e.g., from $^{55}\mathrm{Fe}$ X-ray hits) will be important for assessing these devices' ultimate performance.  %


%
\subsection{Dark current}\label{sec:darkcurrent}
One of the key performance metrics for photon-counting detectors such as skipper CCDs is the rate $\lambda$ of single-electron events generated by non-astrophysical processes, e.g., thermal dark current. The preliminary requirements for \textit{HWO}, for example, set a limit of $\lambda \lesssim 10^{-4}$ e$^-$/pixel/s. We use deep multi-sample dark frames (50 rows $\times$ 600 columns $\times$ 600 samples) to measure the dark current. Since the readout time of each image is 15 minutes, 18 seconds of exposure are accumulated per row. After performing the usual baseline subtraction and gain calibration, we mask all pixels with a threshold of ${\ge}5$ median absolute deviations (MAD) above the row's median, or 15$\mathrm{e}^-$, whichever is greater. We also mask these pixels' neighbors out to a radius of 2 pixels. This ensures diffuse charge from energetic particles does not corrupt the dark-current measurements (though it does not remove some of the low-energy Cherenkov photons emitted from those tracks \cite{Moroni:2025, Gaido:2025}). Finally, we fit the median charge in each row as a function of its exposure time using a linear function $Q(n_\mathrm{row}) = \lambda n_\mathrm{row} \Delta t_\mathrm{row} + C$ to extract $\lambda$ (the constant $C$ represents the exposure-independent spurious charge, e.g., clock-induced charge).

\par We observe differences in the dark-current behavior of the two CCDs tested. For CCD \#1, we found $\lambda \approx (2-3)\times 10^{-4}$ e$^-$/pix/s prior to irradiation, and a baseline of $\approx 5\times 10^{-4}$ e$^-$/pix/s several days after irradiation. For CCD \#2, improved light-tightness of the sensor package resulted in much lower single-photon rates: both prior to irradiation and ${\sim}$90~h post-irradiation, $\lambda \lesssim 1.5\times 10^{-4}$ e$^-$/pix/s for all amplifiers. Similar p-channel skipper CCDs operated at these temperatures in shielded environments (e.g., for dark-matter experiments) have measured dark count rates $\lambda \lesssim 10^{-6}$ e$^-$/pix/s ~\cite{PerezGarcia:2024lqj,SENSEI:2024yyt,DAMIC-M:2025luv}, suggesting that the measured $\lambda$ in CCD \#2 both pre- and post-irradiation is dominated by environmental sources. This is not surprising, since the sensor package had a small hole for LED flat-field illumination (allowing stray light to ingress) and the device readout sequence was not optimized for low clock-induced charge. Thus, our measurements of the dark current should be taken as a practical upper limit given the testing configuration, rather than an indication of the intrinsic performance of a fully-optimized sensor package. We conclude that thick p-channel skipper CCDs operated at 140~K can readily meet the baseline requirement of $\lambda \lesssim 10^{-4}$ e$^-$/pix/s for missions such as HWO even after a 10-year-equivalent DDD. 
\par By studying the dependence of $\lambda$ on CCD temperature, we are able to identify the trap species involved in dark-current generation. The bulk dark current generated in thick fully-depleted CCDs is observed to scale with temperature following~\citep[e.g.,][]{Janesick:2001,2002SPIE.4669..193W,PerezGarcia:2024lqj}
\begin{equation}
    \lambda_\mathrm{bulk}(T) = T^{3/2} \sum_i \lambda_{i}  \exp\left[-\frac{E_{a,i}}{kT}\right]
\end{equation}
where the sum runs over the set of trapping states in the device with temperature-independent coefficients $\lambda_i$ and activation energies $E_{a,i}$ above the valence band. As shown in Fig.~\ref{fig:arrhenius}, we find that two trapping states dominate the dark-current evolution, a shallow trap ${\sim}$0.14~eV above the valence band consistent with the oxygen-vacancy (A-center), and a deep trap ${\sim}$0.56--0.63\,eV above the valence band (i.e., very near to mid-gap $E_g/2\approx0.56\mathrm{\,eV}$), consistent with previous irradiation studies of p-channel non-skipper CCDs~\cite{1039641}. Though we were unable to perform a pre-irradiation temperature sweep with either CCD, it is instructive to compare against the $\lambda(T)$ from a non-irradiated sensor from Vendor \#1 with the same dimensions and operating conditions as our CCD \#1. Those data show a shallow trap consistent with the ${\sim}$0.14-eV state identified in CCD~\#2 (suggesting that measuring CCD \#1 below 160~K would have revealed it there as well), in addition to the ${\sim}$0.56--0.65-eV state found in CCD~\#1.
\par These elevated-temperature tests also enable us to compare our measured $\lambda$ for CCD~\#2 to the universal dark-current damage factor in silicon
\begin{equation}
    K_\mathrm{dark} = \frac{\Delta \lambda(300\mathrm{\,K})}{V_\mathrm{dep} \times \mathrm{DDD}},
\end{equation}
where $\Delta \lambda(\mathrm{300\,K})$ is the radiation-induced increase in dark current as measured at room temperature, $V_\mathrm{dep} \approx (15\times 15\times 650)\,\upmu\mathrm{m}^3$ is the depleted volume of the pixel, and $\mathrm{DDD}\approx 1.1\times 10^7\,\mathrm{MeV/g}$ is the absorbed dose from our 10-year fluence. Since our CCDs are not operable at room temperature, we extrapolate $\Delta \lambda(300\mathrm{\,K})$ using the 160-K dark current measurement and the deep-trap energy level, obtaining\footnote{We compare this result to $\Delta \lambda(\mathrm{300\,K}) = 114 \mathrm{\,pA/cm}^2 \approx 1.6\times 10^{3}$ e$^-$/pix/s obtained using non-irradiated p-channel skipper CCDs with the same pixel areas and thickness~\cite{PerezGarcia:2024lqj}. } $\Delta \lambda (300\mathrm{\,K})\approx 3\times 10^5\mathrm{\,e}^-\mathrm{/pix/s}$. We find $K_\mathrm{dark}\approx 2\times 10^5\mathrm{\,e}^-/\mathrm{cm}^3/\mathrm{s}/(\mathrm{MeV/g})$, in excellent agreement with the universal silicon value $(1.9\pm0.6)\times 10^5 \mathrm{\,e}^-/\mathrm{cm}^3/\mathrm{s}/(\mathrm{MeV/g})$ at 300~K~\citep[e.g.,][]{903792}. These results indicate that the post-irradiation dark current in our CCDs above ${\sim}$160~K is dominated by radiation damage rather than pre-existing fabrication defects. We note that the universal value also assumes a one-week room-temperature anneal following irradiation, while our CCDs remained below ${\sim}$160~K for more than 3 days and were returned to room temperature over the course of several hours (during which the dark-current measurements were taken). Therefore, our value of the damage factor reflects the defect landscape of the silicon before any significant (reverse) annealing had taken place.  

\begin{figure}[htpb]
    \centering
    \includegraphics[width=0.95\linewidth]{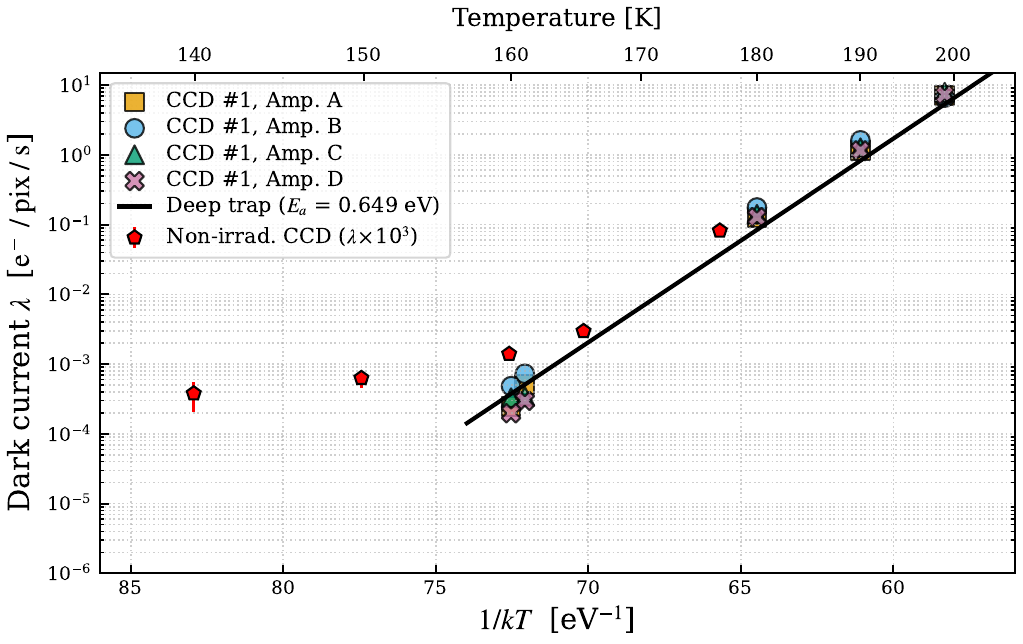}

    \vspace{0.5cm}

    \includegraphics[width=0.95\linewidth]{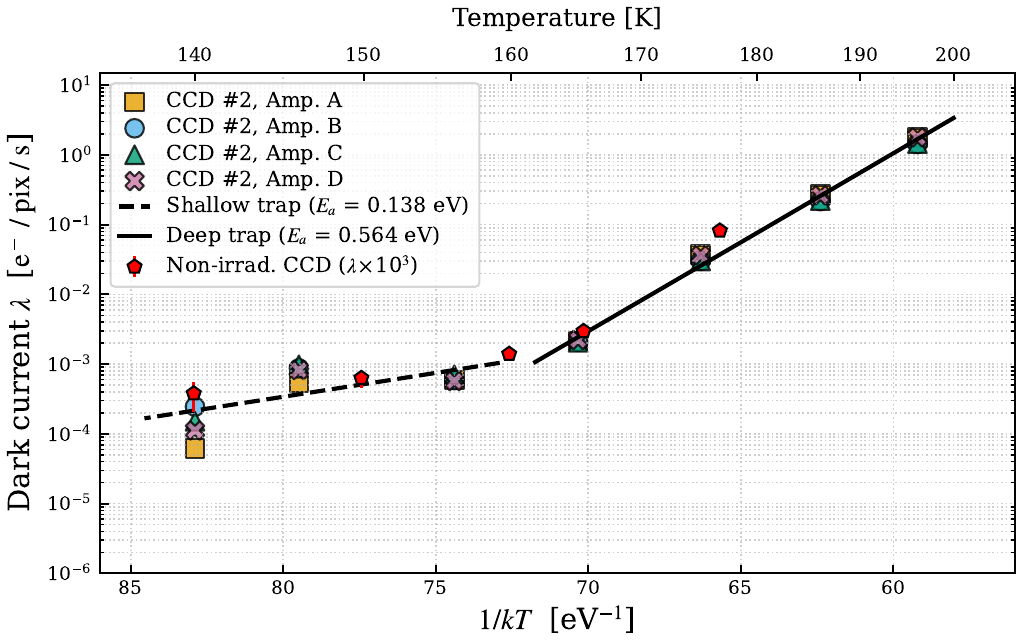}
    \caption{\textbf{(Top)} Arrhenius plot for CCD \#1 post-irradiation, showing the evolution of dark current with temperature and the consistency with a mid-level trap near $E_a \approx 0.63\mathrm{\,eV}$. We also plot measurements from a non-irradiated sensor from Vendor~\#1~\cite{PerezGarcia:2024lqj}, scaled up by a factor of $10^3$ for visibility. \textbf{(Bottom)} Same, but for CCD \#2. The lower initial temperature of 140~K allows us to identify a separate trap species at $E_a \approx 0.14\mathrm{\,eV}$. For more details see Sec.~\ref{sec:darkcurrent}}
    \label{fig:arrhenius}
\end{figure}

\par Finally, we consider the evolution of the dark current over time as shown in Fig.~\ref{fig:activation_time_series} for CCD \#2. (We do not study the time-series data for CCD \#1 owing to its elevated operating temperature.) The data for all four amplifiers are well-described by a double-exponential function, finding component time constants of ${\sim}$5.3~h and ${\sim}$38.3~h. We note that previous tests of conventional p-channel CCDs following proton irradiation also showed double-exponential behavior at a temperature of 133~K~\cite{Dawson:2007yi}, with time constants ${\sim}$60~h and ${\sim}$330~h. Assuming this double-exponential time dependence is the result of two species of radiation-induced traps emptying, and that the same trap species are relevant at 133~K and 140~K, we can express the activation energy difference between the two trap species (A and B) as
\begin{equation}
    E_B - E_A = kT \ln\left(\frac{\tau_A}{\tau_B} \frac{\sigma_B}{\sigma_A}\right).
\end{equation}
Since the energy difference is only logarithmically dependent on $\sigma_B/\sigma_A$, we can safely make the simplifying assumption that $\sigma_A \approx \sigma_B$. Our results at 140~K indicate $E_B - E_A \approx 0.024\mathrm{\,eV}$, and the results of Ref.~\cite{Dawson:2007yi} at 133~K indicate $E_B - E_A \approx 0.020\mathrm{\,eV}$. This strongly suggests that both datasets share the same underlying trap dynamics. To place the energies in absolute terms, we assume a typical cross section $\sigma \sim 10^{-15}\mathrm{\,cm}^2$ and obtain $E_A \approx 0.4\mathrm{\,eV}$ above the valence band. Deep-level transient spectroscopy of n-type silicon has previously identified similar structures which have been attributed to higher-multiplicity vacancy clusters~\citep[e.g.,][]{Kovacevic_2005,cryst12121703}. We propose that these defects formed during the initial proton irradiation, captured charge, and were prevented from annealing into more stable configurations (e.g., $\mathrm{V}_2$) by the low operating temperature. Further irradiation studies at lower temperatures may help to test this hypothesis.

\begin{figure}
    \centering
    \includegraphics[width=0.9\linewidth]{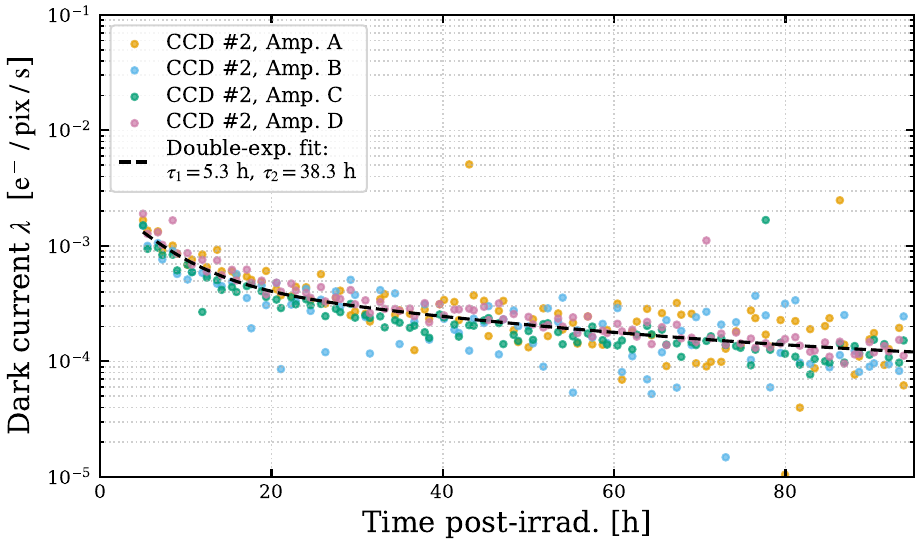}
    \caption{Time-series evolution of the dark current for the four quadrants of CCD~\#2, with measurements beginning ${\sim}$5 h after irradiation. The sensor was maintained at 140~K for the entire time. For more details see Sec.~\ref{sec:darkcurrent}.}
    \label{fig:activation_time_series}
\end{figure}


%
\subsection{Hot Pixels}\label{sec:hotpixels}
High-energy particle bombardment of CCDs is known to produce ``hot'' pixels with excessive dark current, likely as a result of displacement damage in the lattice creating mid-gap charge generation centers~\citep[e.g.,][]{Janesick:2001}. These pixels can bleed charge during readout, resulting in vertical trails that contaminate large fractions of the image. To search for hot pixels, we use the $N_\mathrm{samp} = 1$ dark frames taken pre- and post-irradiation with pixel exposure times ranging from 0--80~s. Following standard bias subtraction and gain calibration, we median-combine the 700 post-irradiation images to filter out high-energy particle tracks. This has the effect of reducing the readout noise of the combined stack to $\sigma_{700} \approx 1.2 \sigma_1 / \sqrt{700} \approx 0.18\mathrm{\,e}^-\mathrm{/pix\;(rms)}$, where the factor of 1.2 arises from the use of the median (rather than the mean) to estimate the rms. After searching for persistent pixels with raw charge ${>}5\sigma_{700}$ and correcting for the position-dependent pixel exposure, we find the results shown in Fig.~\ref{fig:hotpixels}. (We do not show CCD ~\#1 since the thermal dark current at its 160-K operating temperature prevents identification of low-charge pixels.) These results are consistent with pre-irradiation measurements, though the latter only consisted of ${\sim}$5 frames with the same exposure and thus the hot-pixel threshold is higher. In particular, we note that following a 10-year-equivalent DDD at Earth/Sun L2, the fraction of hot pixels emitting at least 1 e$^-$/pix/min was $\lesssim 10^{-4}$. Though dark-current rates in silicon (particularly for lattice defects) are extremely temperature-sensitive\footnote{We note that focal-plane temperatures well below 130--140 K are achievable at L2 with a combination of passive radiators and sun-shielding~\citep[e.g.,][]{2016JATIS...2d1212R,Gardner_2023,2025A&A...697A...1E}. While operating n-channel CCDs at ${\sim}$130--140~K would suppress the hot-pixel rate, it would likely significantly worsen the CTI from A-center and E-center traps by increasing their $\tau_e$.}, it is instructive to compare several space-telescope missions at their respective operating temperatures. The n-channel CCDs on \textit{HST} run significantly warmer (190 K), which significantly enhances the rate of hot pixels. On the other end of the scale, EMCCD201-20 devices similar to those which will fly on \textit{Roman} at L2 were tested at significantly colder temperatures (${\sim}$165 K), with pre-launch tests finding ${\lesssim}0.01\%$ of pixels above a 3 e$^-$/pix/min threshold~\cite{bush_thesis}. Comparing to previous beam tests of conventional p-channel CCDs~\cite{Dawson:2007yi}, we find that our skipper CCD hot-pixel rates are an order of magnitude lower. Since our delivered DDD was an order of magnitude lower and our temperature was 7 K higher than Ref.~\cite{Dawson:2007yi}, we would expect comparable hot-pixel rates. This is likely a result of improvements in the manufacturing process to reduce trace impurities in the silicon wafers (particularly carbon and oxygen), and further demonstrates the robustness of modern high-resistivity p-channel CCDs.
\begin{figure}
    \centering
    \includegraphics[width=0.95\linewidth]{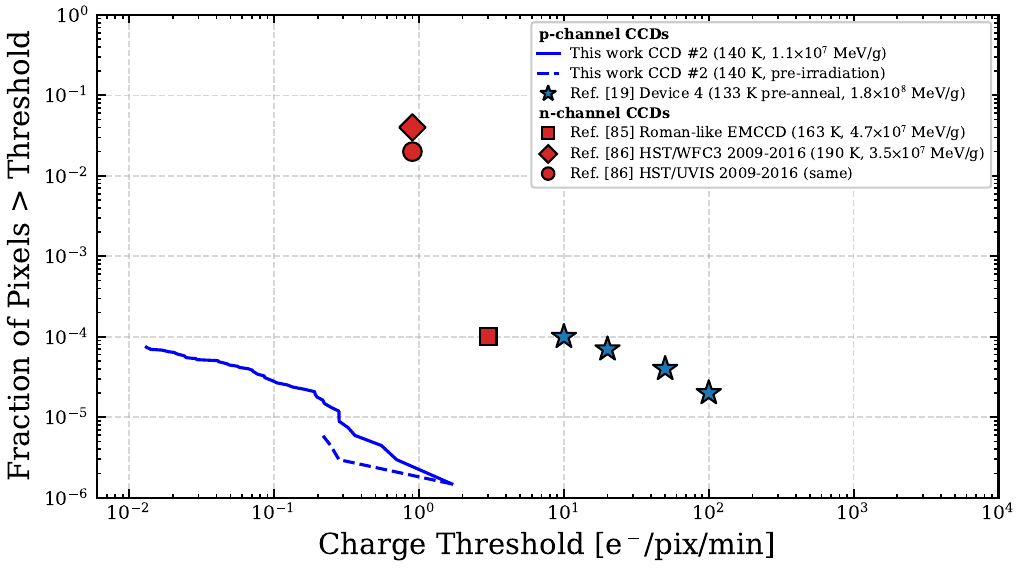}
    \caption{Fraction of hot pixels (with charge exceeding a given cutoff) for skipper CCD~\#2 (blue line) following its maximum proton dose. For context, we also include results from conventional p-channel CCDs~\cite{Dawson:2007yi} as well as instruments on \textit{HST}~\cite{bourque2016wfc3} and \textit{Roman}~\cite{bush_thesis}. The average DDD rate in low-Earth orbit (\textit{HST}) is ${\sim}5.2\times 10^6\mathrm{\,MeV/g/yr}$~\citep[e.g.,][]{jones2000acs}, and at L2 (\textit{Roman/HWO}) is ${\sim}1.1\times 10^7\mathrm{\,MeV/g/yr}$ (Sec.~\ref{sec:rad_damage_calcs}). No attempt has been made to correct for differences in operating temperature or NIEL dose. For more details see Sec.~\ref{sec:hotpixels}.}
    \label{fig:hotpixels}
\end{figure}
\subsection{Output Transistor Curves}\label{sec:transistors}
As described in Sec.~\ref{sec:campaign2}, we used an external power supply as well as the LTA to sweep through a range of $V_\mathrm{ref}$ and $V_{dd}$, measuring the output voltage $V_s = V_\mathrm{video}$ with a digital multimeter. The applied substrate voltage was 70 V. For convenience, we define the output current 
\begin{equation}\label{eq:ids}
    I_{ds} = -V_s/R_L
\end{equation}
where $R_L = 20\mathrm{\,k}\Omega$ is the load on the video line. We also define the threshold voltage $V_t$ of the gate MOSFET to be the value of $V_{gs}$ where $I_{ds} = 0$ (at $V_{ds} = -7\mathrm{\,V}$), which we obtained by numerical interpolation of the transistor curves. 
\par The results of the transistor-curve scan for CCD \#2 are shown in Fig.~\ref{fig:transistors} and Table~\ref{tab:transistors}. The consistency between pre- and post-irradiation behavior in the subthreshold regions (particularly in the values of $V_t$) indicates that TID effects are minimal, which is to be expected since the absorbed dose was ${\ll}$1~krad. The behavior of the transistor curves in the cutoff region above $V_t$, however, deserves additional attention. Prior to irradiation, amplifiers A and (to a lesser extent) B display subthreshold humps, indicative of minor process variations leading to parasitic edge conduction. For $V_{gs} \gtrsim 13.5\mathrm{\,V}$, all four amplifiers display\footnote{We note that Fig. 9 of Ref.~\cite{Cervantes-Vergara:2025xis} (obtained with a different sensor from the same vendor) shows the same linear behavior in the cutoff region, further supporting that it is a test-stand artifact rather than a sensor characteristic.} linear/Ohmic $I-V$ characteristics. After irradiation, however, all four output transistors exhibited Ohmic behavior for $V_{gs} \gtrsim \text{12.8--13\,V}$. Since this region exhibits simiular behavior pre- and post-irradiation, we attribute it to a parasitic conduction path (likely in the outside-cryostat breakout board used to probe the voltages) rather than a radiation-induced defect in the silicon itself. In particular, Eqn.~\eqref{eq:ids} shows that any stray voltage on the $V_\mathrm{video}$ probe is immediately converted to a current $I_\mathrm{ds}$. For example, we observe $V_\mathrm{video} \approx 70\mathrm{\,mV}$ at $V_\mathrm{ref} = 14\mathrm{\,V}$, irrespective of the value of $V_{dd}$. This is consistent with there being a parasitic conduction path with $R_p \approx 3\mathrm{\,M}\Omega$ between $V_\mathrm{ref}$ and $V_\mathrm{video}$ traces in the warm electronics chain (e.g., from surface residue or moisture on the PCB). Further irradiation campaigns optimized for total ionizing dose (e.g., at electron-beam or gamma-ray facilities) will likely be necessary to fully ascertain the ionizing-dose hardness of the output transistor.

\begin{table}[t]
    \centering
    \begin{tabular}{ccc}
    \toprule
        \textbf{Amplifier} & $\boldsymbol{V_t}$ \textbf{pre-irrad. [V]} & $\boldsymbol{V_t}$ \textbf{post-irrad. [V]} \\
        \hline
        A & 12.95 & 12.82 \\
        B & 12.81 & 12.85 \\
        C & 12.60 & 12.60 \\
        D & 12.79 & 12.73 \\
        \bottomrule
    \end{tabular}
    \caption{Threshold voltages for CCD \#2 measured before and after irradiation, setting $V_{ds} = -7\mathrm{\,V}$. Uncertainties are at the $\pm$10 mV level, dominated by the numerical interpolation and the resolution of the voltmeter.}
    \label{tab:transistors}
\end{table}

\begin{figure}[H]
    \centering
    \includegraphics[width=0.8\linewidth]{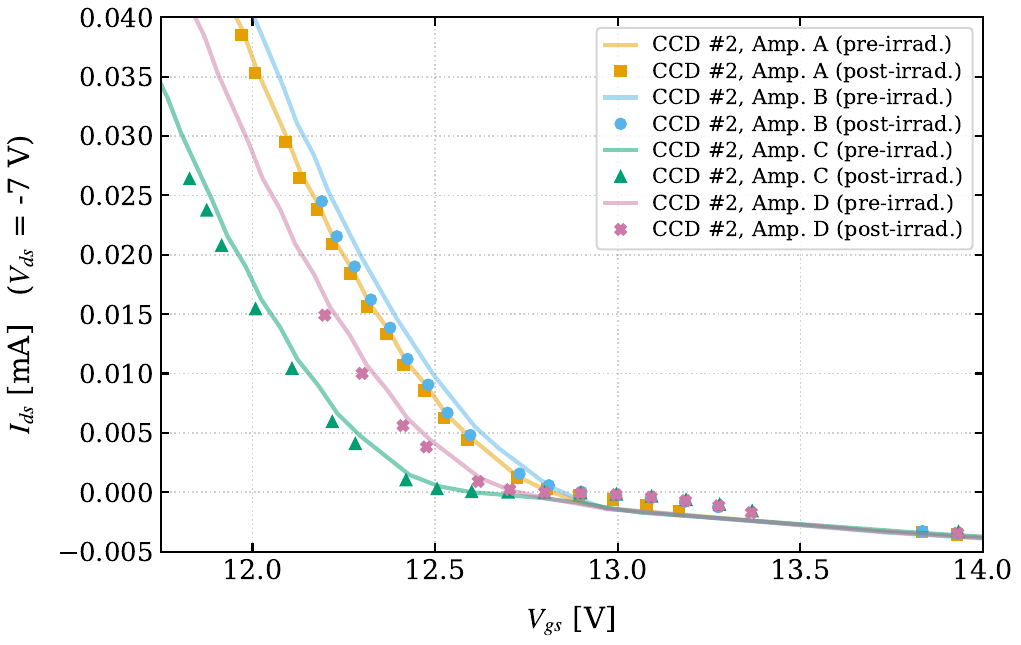}

    \caption{Output transistor curve behavior for CCD \#2 at 140~K before (solid lines) and after (points) proton irradiation, showing expected $I-V$ characteristics above $V_t$. The negative linear trend in the cutoff region is attributed to a parasitic conduction path in the warm electronics chain. For more details see Sec.~\ref{sec:transistors}.}
    \label{fig:transistors}
\end{figure}

\section{Implications for Space-Based Instrumentation}\label{sec:discussion}
\par From the previous discussion, it may appear that p-channel CCDs are protected from the effects of displacement damage, both on account of their majority carriers being holes and their low operating temperatures. The latter factor, however, may present significant challenges for spacecraft operations. Though our results show that p-channel skipper CCDs operating near ${\sim}$140~K are mostly protected from dark current and hot-pixel generation, many of the underlying vacancies and interstitials do not disappear or form stable defects; rather, they simply remain frozen in the lattice until sufficient thermal energy is supplied to move them~\citep[e.g.,][]{PhysRevB.59.3969,Newman_1982,WATKINS2000227,KOVACEVIC2005223}. If the focal-plane temperature increases above the defect-migration barrier (e.g., ${\sim}$0.45~eV for the $\mathrm{V}^0$ common in the depletion region~\cite{10.1063/1.2937198}), the vacancies rapidly become mobile again, forming stable defects such as $\mathrm{V}_2$ and $\mathrm{V}_n \mathrm{O}_m$ at temperatures ${\gtrsim}$200~K. In contrast, charged vacancies common in the highly-doped n- or p-channels have significantly lower migration barriers, becoming mobile at temperatures ${\gtrsim}$80~K and ${\gtrsim}$160~K, respectively~\citep[e.g.,][]{BOURGOIN1972135,Newman_1982}. This is the well-known reverse annealing phenomenon, in which both n- and p-channel CCDs may experience permanent degradation in CTI and dark current following thermal cycles even after the device is returned to low temperatures~\citep[e.g.,][]{2005ITNS...52..519B,Grant:2008tk,MONMEYRAN201623}. In particular, trap-pumping studies of p-channel CCDs have found significant differences in the defect populations depending on the thermal history of the sensor~\citep[e.g.,][]{8048486}. Devices irradiated at room temperature display large concentrations of $\mathrm{V}_2$; in contrast, devices irradiated at ${\sim}$150~K showed ${\sim}1/3$ the number of identified defects, and the approximately flat distribution of emission time constants spanned several orders of magnitude. Crucially, devices irradiated cold and then subjected to room-temperature annealing stages displayed the both the main $\mathrm{V}_2$ peak and as well as an additional defect species attributed to interstitial carbon (C$_i$). Furthermore, the C$_i$ contribution was not observed in either the room-temperature-irradiated device or the device irradiated and tested at cryogenic temperatures, suggesting that the annealing phase itself was responsible. 
\par These facts may present significant constraints on spacecraft operations, particularly for telescopes which require periodic bake-outs to clear volatiles from the optical elements~\citep{10.1117/12.552029}, or during trap-pumping cycles to characterize the properties of lattice defects. We emphasize that this is not a consequence of particle fluence incurred while the CCD is warm (though this may also be important during the post-launch/commissioning phase); rather, it is the thermal activation of latent damage accumulated while irradiated cold. While p-channel devices may be protected from most of the mid-level dark-current-generating traps on account of their low operating temperature, any stable defects with $\tau_e \gtrsim \Delta t$ will permanently degrade the CTI. Additionally, the low operating temperature of p-channel CCDs tends to make $\tau_e$ quite long; depending on the defect involved, it may be longer than the image readout time, making it impossible to link CTI trails to their source as has been done with \textit{HST}}~\citep[e.g.,][]{2026MNRAS.546f2186M}. Further work will be necessary to determine the lifetimes of defects in devices irradiated and maintained at cryogenic temperatures for significant periods of time, as well as the temperatures at which the latent lattice damage becomes mobile. It may also be possible to adapt ``defect engineering'' techniques from the high-energy physics community, doping CCDs with elements which would lock up interstitials and vacancies into more electrically benign complexes~\citep[e.g.,][]{LUUKKA2004152}.

\section{Conclusions}\label{sec:conclusions}
\par Fully depleted p-channel CCDs are an extremely attractive candidate for focal-plane sensors in space instrumentation, owing to their enhanced radiation hardness compared to their n-channel counterparts. Our results provide the first conclusive evidence that this p-channel advantage applies to photon-counting skipper CCDs as well, demonstrating for the first time their radiation-hardness in realistic thermal and electrical environments. By subjecting two sensors from different vendors to 10-year-equivalent DDDs without intervening thermal annealing, we show that p-channel skipper CCDs maintain low readout noise, CTI, dark current, and hot pixel fractions at their nominal operating temperatures of 140--160~K. We note that our results in this study concentrate on displacement damage rather than total ionizing dose; therefore, future study will be needed to determine the TID hardness of these sensors and their readout systems. Our results demonstrate that p-channel skipper CCDs are remarkably resilient to displacement damage in representative thermal environments; however, future experimental efforts are needed to assess these sensors' TID resilience under the conditions expected in deep space. We conclude by noting that we also irradiated several other silicon-based photon-counting sensors such as multi-amplifier sensing (MAS) CCDs~\citep[e.g.,][]{Lin_2024,Botti_2024} and single-electron sensitive readout (SiSeRO) CCDs~\citep[e.g.,][]{Chattopadhyay:2021xiu,PhysRevLett.133.121003}, and characterization results are forthcoming. These sensors are based on similar p-channel CCD architecture as skipper CCDs but are optimized for significantly faster readout rates while maintaining sub-electron readout noise. If these devices maintain the same radiation-hardness as previous p-channel devices, they will likely be extremely attractive candidates for space instrumentation.

\section*{Acknowledgements}
We are grateful to the staff at the Northwestern Medicine Proton Center for their assistance with this testing, especially Steve Laub and the rest of the Physics department. We also thank Jim Hirschauer for assistance and coordination at the early stages of planning, and our CCD colleagues at Fermilab (particularly Michelle Jones and Andy Lathrop) for their assistance in detector packaging and preparation. The fully depleted skipper CCD was developed at Lawrence Berkeley National Laboratory, as were the designs described in this work. This work was partially supported by NASA APRA award No.\ 80NSSC22K1411 and a grant from the Heising-Simons Foundation (\#2023-4611). BR was partially supported by a KICP Fellowship at the University of Chicago.  Fermilab is managed by Fermi Forward Discovery Group, LLC, acting under Contract No. 89243024CSC000002 for the U.S. Department of Energy. A shortened version of this manuscript is being prepared for submission to the 2026 SPIE Astronomical Telescopes+Instrumentation conference as a proceedings; the present manuscript reflects the expanded version of record and will be updated with a DOI to the proceedings when available.
\appendix
\section{Tables}
\begin{table}[h!]
    \centering
    \caption{Readout noise and gain for all four output amplifiers of the p-channel LBNL CCDs with $N_\mathrm{samp}=300$. For each operating condition, the top row shows the noise $\sigma_1$ [e$^-$ rms/pixel], and the bottom row shows the gain $g_e$ [ADU/e$^-$]. Uncertainties on $\sigma_1$, $g_e$, and $\alpha$ are approximately $\pm 2\%$, $\pm 5\%$, and $\pm 10\%$, respectively. }
    \label{tab:ccd_amps_detailed}
    \renewcommand{\arraystretch}{1.2} 
    \setlength{\tabcolsep}{8pt} 
    \begin{tabular}{llccccc}
        \toprule
        \multirow{2}{*}{\textbf{Device}} & \multirow{2}{*}{\textbf{Condition}} & \multirow{2}{*}{\textbf{Metric}} & \multicolumn{4}{c}{\textbf{Amplifier}} \\ 
        \cmidrule(lr){4-7}
        & & & \textbf{A} & \textbf{B} & \textbf{C} & \textbf{D} \\
        \midrule
        
        \multirow{12}{*}{\textbf{CCD \#1}} 
            & \multirow{2}{*}{Pre-Irradiation} 
                & $\sigma_1$ & 4.12 & 4.82 & 3.94 & 4.12 \\
            &   & $g_e$      & 84.0 & 68.9 & 84.0 & 86.6 \\

            \addlinespace 
            
            & \multirow{2}{*}{Irradiation I}   
                & $\sigma_1$ & 4.20 & 4.62 & 3.97 & 3.90 \\
            &   & $g_e$      & 82.8 & 70.6 & 83.9 & 85.2 \\

            \addlinespace

            & \multirow{2}{*}{Irradiation II}  
                & $\sigma_1$ & 4.01 & 4.65 & 4.01 & 3.91 \\
            &   & $g_e$      & 82.9 & 71.1 & 81.8 & 85.7 \\
            \addlinespace

            & \multirow{2}{*}{Irradiation III} 
                & $\sigma_1$ & 4.32 & 4.83 & 4.43 & 3.83 \\
            &   & $g_e$      & 82.5 & 70.2 & 83.3 & 86.0 \\

            \addlinespace
            
            & \multirow{2}{*}{Irradiation IV}  
                & $\sigma_1$ & 3.94 & 4.83 & 4.36 & 4.21 \\
            &   & $g_e$      & 83.4 & 70.8 & 84.0 & 83.0 \\

            \addlinespace

            & \multirow{2}{*}{Long-term Base.} 
                & $\sigma_1$ & 4.08 & 4.90 & 4.29 & 3.78 \\
            &   & $g_e$      & 83.4 & 67.5 & 83.6 & 85.5 \\

        \midrule
        
        \multirow{7}{*}{\textbf{CCD \#2}} 
            & \multirow{2}{*}{Pre-Irradiation} 
                & $\sigma_1$ & 3.50 & 3.78 & 3.53 & 3.47 \\
            &   & $g_e$      & 93.1 & 94.3 & 94.8 & 94.6 \\

            \addlinespace
            
            & \multirow{2}{*}{Irradiation I}   
                & $\sigma_1$ & 3.56 & 4.04 & 3.66 & 3.56 \\
            &   & $g_e$      & 90.3 & 85.9 & 90.3 & 91.4 \\

            \addlinespace
            
            & \multirow{2}{*}{Long-term Base.} 
                & $\sigma_1$ & 3.44 & 3.43 & 3.45 & 3.47 \\
            &   & $g_e$      & 91.7 & 91.4 & 95.6 & 94.5 \\

        \bottomrule
    \end{tabular}
\end{table}
\pagebreak

\bibliography{biblio}
\bibliographystyle{spiebib}

\end{document}